\documentclass[bibyear]{aa} 

\usepackage{graphicx}
\usepackage{txfonts}
 \usepackage{natbib} 
 \usepackage{subcaption}
\usepackage{hyperref}
\usepackage{comment}
\usepackage{amsmath}      
\usepackage{amssymb}      
\usepackage{mathtools}
\usepackage{dblfloatfix}
\usepackage{booktabs}

\bibpunct{(}{)}{;}{a}{}{,}
\defcitealias{Dodd_2023}{D23}

\begin{document}

   \title{Two faces of Gaia-Sausage-Enceladus}

   \subtitle{Mining the chemical abundance space with graph attention networks}

   \author{
      Milan Quandt-Rodriguez\inst{1,2}\thanks{\email{milan.quandt@inaf.it}} 
      \and
      Sara Lucatello\inst{1} 
      \and
      Lorenzo Spina\inst{1,3} 
      \and
      Mario Pasquato\inst{5} 
      \and
      Marco Canducci\inst{6}
      }

   \institute{ INAF - Osservatorio Astronomico di Padova, Vicolo dell’Osservatorio 5, 35122 Padova, Italy
         \and
             Dipartimento di Fisica e Astronomia, Universit\`a di Padova, I-35122, Padova, Italy
        \and
             INAF - Osservatorio Astrofisico di Arcetri, Largo E. Fermi 5, 50125, Firenze, Italy            
        \and
             INAF - Osservatorio Astronomico di Brera, via Brera 28, 20121  Milano, Italy
        \and 
             INAF - IASF-Milano, Via Alfonso Corti 12, I-20133 Milano, Italy
        \and 
             University of Birmingham, School of computer science, B15 2TT, Birmingham, United Kingdom
             }

   \date{Received November 7, 2025; accepted January 29, 2026}

% \abstract{}{}{}{}{} 
% 5 {} token are mandatory
 
  \abstract
   {}
   {Recent studies suggest that chemical abundances hold the key to disentangling halo substructure, providing a more reliable tracer than dynamics alone. We aim to probe the Milky Way stellar halo using high-dimensional chemical abundances from GALAH DR4.  By leveraging multiple nucleosynthesis channels in synergy with integrals of motion (IoM), we extract information hidden in the raw abundance space to perform chemical tagging.}
   {With a graph attention autoencoder, we reconstruct a dynamics-informed, denoised chemical space and identify coherent stellar substructures by applying ensemble clustering.}
   {Our method successfully recovers the three largest globular clusters hidden in the dataset, estimates the in-situ fraction to be approximately 41\%, and chemically characterizes several dynamical halo substructures. Strikingly, stars dynamically associated with Gaia-Sausage-Enceladus (GSE) separate into two chemically distinct clusters. By examining their abundances, energy ($E$) and angular momentum ($L_z$) distributions, together with the metallicity trend with $E$, we connect these clusters to their birthplace within the progenitor by proposing a simple infall scenario: one cluster traces the metal-poor, less evolved outskirts, while the other traces the metal-rich, chemically evolved core. }
   {These findings support the picture of GSE as a massive disk galaxy with a noticeable negative metallicity gradient. Moreover, selecting potential GSE members based on their chemical properties rather than solely on their dynamics naturally expands the membership to stars spanning a wider range of IoM.}


   \maketitle
%
%-------------------------------------------------------------------

\section{Introduction}
Galaxy mergers play a key role in the hierarchical cosmological paradigm. They are the primary mechanism by which galaxies acquire the bulk of their dynamical mass \citep{wang}, and thus tracking these mergers is crucial for unraveling the assembly of galactic systems. To track past mergers across long timescales, we must rely on stars. In this context, the stellar halo of the Milky Way (MW) is regarded as the primary tracer of accreted stellar systems: it is here that stars from these disrupted systems are more likely to be deposited \citep{bullock_2005}. Moreover, the halo is also expected to host stars formed within the MW ("in-situ"), originally part of the disk but displaced onto hotter orbits during mergers \citep{tissera, belokurov_2019}. \par

The association of stars with different merger events in the Galaxy is, however, a non-trivial task. Accreted stars lose their spatial coherence after a few gigayears, mixing with each other and with the in-situ component, which makes it impossible to trace them as over-densities in the sky. One approach to overcome this challenge is to rely on dynamics. As several studies have shown \citep{helmi,knebe}, the integrals of motion (IoM) of orbits are conserved to first order during galaxy mergers. These include the energy $E$, vertical component of angular momentum $L_z$, the perpendicular component $L_\perp$ or the associated actions $J_R$, $J_{\Phi}$, and $J_z$ in an axisymmetric potential. This method has enabled important discoveries of relatively recent mergers—such as the Helmi streams \citep{helmi}—and has shown that the local halo is highly structured in this space \citep{Lovdal_2022, Ruiz_Lara_2022, Dodd_2023}, revealing a plethora of candidate accreted substructures. However, it also relies on several assumptions.

Recent studies have shown that when these assumptions are relaxed, the previous approach can no longer unequivocally identify merger debris \citep{jeanbaptiste, amarante_2022, khoperskov, mori}. In particular, \citet{jeanbaptiste} show that when modeling Galactic mergers including kinematical heating, tidal effects, and especially dynamical friction, a single satellite can be responsible for several independent over-densities. Furthermore, over-densities of different satellites can overlap, and even in-situ stars can form substructures in response to the accretion events, raising doubts about whether the various over-densities truly trace distinct accreted systems. Additionally, \citet{mori} show that if the accreted satellite had a radial or vertical metallicity gradient, even the metallicity distribution function  of different clumps in IoM space stops being a strong diagnostic to discern between different progenitors. \par
A more robust approach to the problem is to use chemical abundances. The photospheric composition of low-mass stars remains largely unchanged over a star’s lifetime, with a few exceptions: C and N are affected by the first "dredge-up", and Li is known to deplete over time. Consequently, a star’s photospheric abundances can serve as a tag linking it back to the molecular cloud from which it formed. If different stellar systems have had distinct chemical enrichment histories, this will be reflected in their stars’ photospheric compositions. This method is known as strong chemical tagging \citep{freeman, desilva} and has been widely applied to reconstruct the Galaxy’s merger history. One of the most striking successes of chemical tagging was \citet{nissenshuster}'s identification of a bimodality in [$\alpha$/Fe] vs [Fe/H] among halo stars, later recognized as the signature of the Galaxy’s most massive merger, Gaia–Sausage–Enceladus (GSE) \citep{belokurov_2018, helmi_gse, gallar_gse}. More recently, several studies have employed the [Mg/Mn] versus [Al/Fe] plane to discriminate between accreted and in-situ stellar populations \citep{hawkins_2015, das, fernandes_2022, horta}. Other elements with potentially high discriminative power include s-process elements synthesized in asymptotic giant branch (AGB) star envelopes (e.g., Ba, Y, Sr) \citep{vicenzo_2019}, r-process elements potentially produced in neutron star mergers (e.g., Eu) and elements with both s- and r-process contributions(e.g., Nd) \citep{aguado_2021, rprocess_2021, naidu_2022, ou_2024}.\par

Past spectroscopic surveys such as APOGEE \citep{apogee}, GALAH \citep{buder19}, H3 \citep{H3}, and Gaia-ESO \citep{gaiaeso}, combined with ongoing surveys such as LAMOST \citep{lamost}, SDSSV Milky Way Mapper \citep{sdss}, DESI \citep{Cooper_2023}, and the upcoming 4MOST \citep{4most} and WEAVE \citep{weave}, provide—or will provide—the chemical abundance data needed for chemical tagging of halo stars.
 However, performing chemical tagging directly using density-based clustering algorithms on raw abundances, even on high quality spectroscopic data, has proven extremely challenging \citep{Spina_2022}. This does not necessarily indicate a lack of structure or information in the dataset; rather, it reflects both methodological limitations and intrinsic constraints imposed by overlapping chemical signatures of stars from different birth sites \citep{Casamiquela_2021}. Factors such as the curse of dimensionality, increasing measurement errors at low metallicity, error bars that are comparable to the intrinsic differences expected in some environments (even at high metallicity), intrinsic abundance scatter between different progenitor systems, and the highly uneven number of stars contributed by distinct substructures all hinder the ability of standard density-based algorithms to detect meaningful stellar groups. To overcome this, several studies have employed Gaussian Mixture Models (GMMs) to partition the chemical-abundance space of halo-star samples into $n$ clusters, subsequently analyzing each cluster independently \citep{das, mackereth, carrillo, buckley}. These approaches, however, require specifying the \textit{a priori} unknown number of clusters, complicating interpretation. Alternatively, some studies apply hard cuts in IoM space to isolate presumed accretion events and then characterize their chemistry \textit{a posteriori} \citep{Naidu_2020, horta}. Yet, recent works show strong overlap between in-situ stars and GSE debris across the entire $E$–$L_z$ plane leading to heavy contamination: \citet{thomas2025}, using the algorithm of \citet{Lovdal_2022, Ruiz_Lara_2022, Dodd_2023} on AURIGA simulations \citep{10.1093/mnras/stx071}, showed that recent mergers (within the last 6–7 Gyr) are reliably recovered, whereas older mergers are often missed. This represents a critical limitation, since the MW’s major mergers concluded approximately 7–8 Gyr ago \citep{hammer_2007, ruchti_2015, carlesi_2020}. Moreover, \citet{mori2025chemicaldecodingkinematicsubstructures} found that contamination in $E$–$L_z$ arises not only from GSE debris but also from the metal-poor disk, further blurring population boundaries. \par

These considerations not only highlight the complexity of the task, but also underscore the need for new methods to detect halo substructure that are guided primarily by chemical abundances, with dynamical information playing a supportive role. The guiding question of this work is: can we fully exploit the rich spectroscopic datasets from large surveys, together with the precise astrometry from Gaia, to achieve robust chemical tagging of halo substructures?

The works by \citet{berni} and \citet{spina_2025} have offered a framework based on graph neural networks (GNNs) to address this very question. At its core, these approaches rest on the idea that a graph structure provides the most suitable representation of the data: chemical abundances are stored as node features, while information about dynamics or orbital properties is encoded in the edges, thereby capturing the underlying topological relationships. Motivated by these results, in this paper we present a novel GNN-based framework for chemically tagging halo stars, which leverages chemical abundance space as the primary guide while incorporating IoM in a complementary, supportive role.
%--------------------------------------------------------------------
\section{Methods}

In many real-world datasets, the entities of interest are not independent but connected through various types of relationships. While traditional approaches focus on individual feature vectors, graph-based representations explicitly encode these connections between "nodes" through "edges". Considering this network structure can reveal collective patterns and dependencies that would remain hidden in feature-based analyses alone \citep{wang_2020}. In the context of halo substructure, chemical abundances already provide valuable insights into common origins, but incorporating dynamical similarity via edges can further enhance the identification of coherent structures. In this work, each node represents an individual star, and each edge encodes dynamical similarity. Additionally, each node stores node features, which are the chemical abundances. Our methodology closely follows \citet{spina_2025}, which we revisit in this section: we first pre-process the original abundance space using a GNN and then apply clustering to the resulting embeddings.

\subsection{Graph neural networks}

\begin{figure}[h!]
    \centering
    \includegraphics[width=\columnwidth]{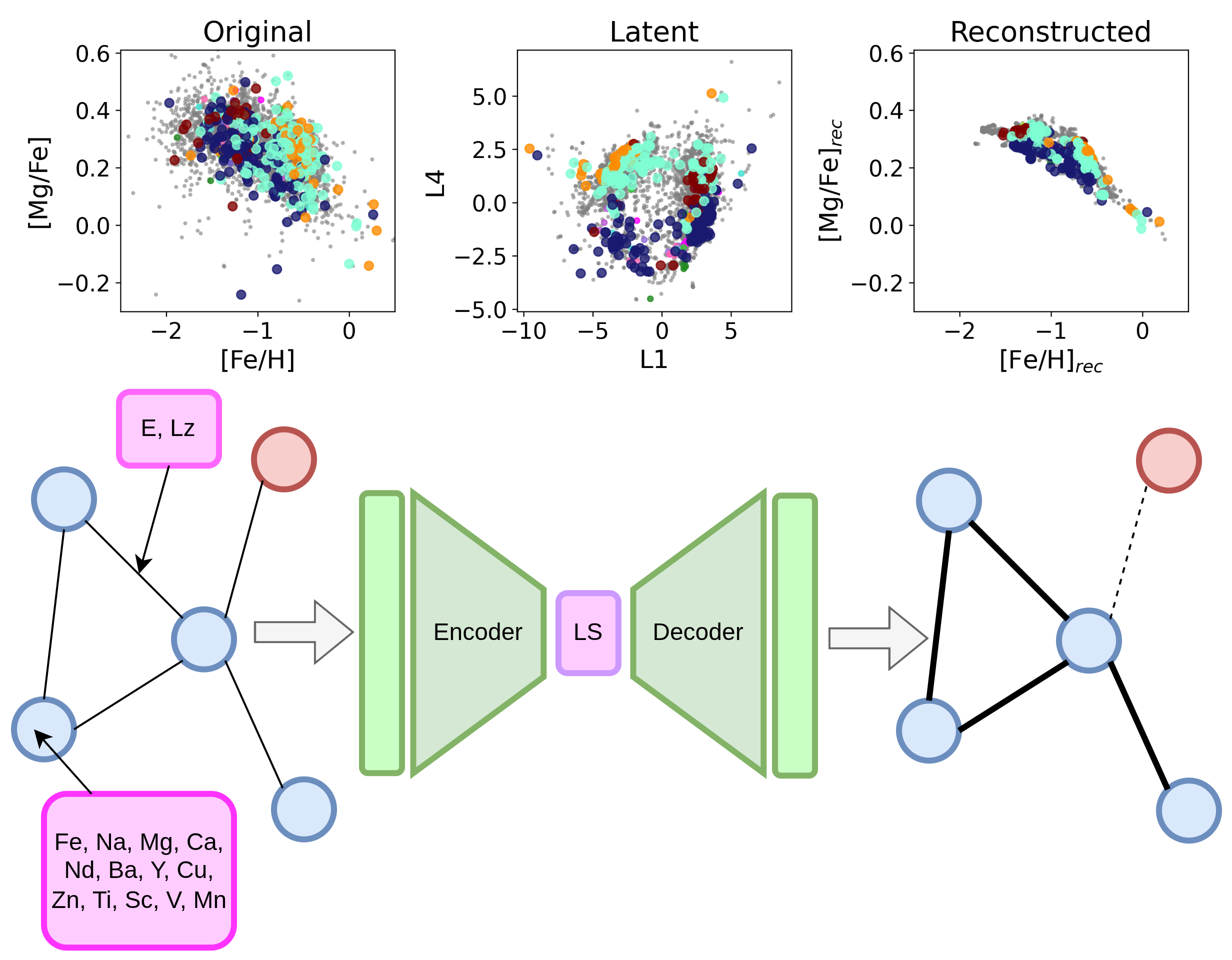}
    \caption{Schematic of the algorithm. Chemical abundances define node features and IoM the edges. The GAT autoencoder reconstructs the node features, learning which links to retain or break. Chemically similar stars remain connected, while dissimilar ones disconnect. On top, original (left) vs. reconstructed abundances (right) with two of the five latent-space dimensions in the middle. This diagram was created with \texttt{draw.io} \citep{drawio}.}

    \label{fig:GAT}
\end{figure}

GNNs were first introduced by \citet{gnn_first} and are a class of NNs designed to operate directly on graph-structured data. Their core principle is message passing: each node updates its representation by aggregating feature information from its neighbors, typically through learnable transformations. Repeated iterations of this process allow information to propagate across the graph, enabling the model to capture both local and global topological structure. In the context of this work, it allows stars to "peek" into the chemical abundances of their dynamical neighbors, enabling each star’s representation to incorporate information about the broader structure. \par 
Mathematically, the embedding of a given node $v$ in a GNN can be written as:
\begin{equation}
h_v^{(k)} = \sigma \Bigg( W \cdot \Big( h_v^{(k-1)} + \operatorname{AGGREGATE} \big( \{ h_u^{(k-1)} : u \in \mathcal{N}(v) \} \big) \Big) \Bigg)
\end{equation}
where $k$ is the layer number, $W$ is the learnable weight matrix, $\sigma$ is a non-linear activation function (e.g., ReLU), $\operatorname{AGGREGATE}$ is a permutation-invariant function (e.g., sum or mean), and $\mathcal{N}(v)$ is the set of neighbors of node $v$. If the aggregation term is omitted, the layer reduces to a standard feedforward NN applied independently to each node. As the GNN has multiple layers, each node’s embedding at layer $k$ incorporates information from its $k$-hop neighborhood, progressively capturing broader structure as $k$ increases.

\subsection{Graph attention networks}

However, this approach is not ideal in our astrophysical context. If two stars are linked solely based on proximity in IoM space, they will still exchange information during message passing even if their chemical compositions differ, because the graph topology is fixed. Ideally, we want a network with a flexible topology, in which the strength of interaction between stars depends only on their chemistry. In this way, chemically similar stars can attend more strongly to each other, while chemically distinct stars can effectively ignore one another, allowing the embeddings to remain grounded in chemistry. \par

This is exactly what graph attention networks (GATs) \citep{gats} are designed to address. They follow the same message-passing scheme described in the previous subsection, but with an important enhancement: each edge is assigned a learnable attention coefficient that determines how much influence a neighbor has on a node’s updated embedding. Mathematically, this is expressed as:
\begin{equation}
h_v^{(k)} = \sigma \Bigg( \sum_{u \in \mathcal{N}(v)} \alpha_{vu}^{(k)} \, W \, h_u^{(k-1)} \Bigg)
\end{equation}

where $\alpha_{vu}^{(k)}$ are the attention coefficients for node $v$ and its neighbor $u$ at layer $k$. The raw attention coefficients are: 
\begin{equation}
e_{vu}^{(k)} = \text{LeakyReLU}\Big(a^\top [W\,h_v^{(k-1)} \, \| \,W\, h_u^{(k-1)}]\Big)
\end{equation}
where \(h_v^{(k-1)}\) and \(h_u^{(k-1)}\) are the embeddings of nodes \(v\) and \(u\) from the previous layer, \(\|\) denotes concatenation, and \(a\) is a learnable vector that transforms the concatenated features into a scalar attention score. The final attention coefficients are obtained by softmax normalization: 
\begin{equation}
    \alpha_{vu}^{(k)} = \frac{\exp(e_{vu}^{(k)})}{\sum_{j \in \mathcal{N}(v)} \exp(e_{vj}^{(k)})}
\end{equation}
where the sum runs over all neighbors \(j\) of node \(v\), ensuring that the attention coefficients \(\alpha_{vu}^{(k)}\) for each node sum to 1.

The embedding of node $v$ itself is included in the sum via a self-loop, crucially allowing a node’s own representation to contribute to its updated embedding.

\subsection{Model architecture and training}

We used an autoencoder architecture, which consists of an encoder that maps input data to a lower-dimensional embedding, called latent space, and a decoder that reconstructs the input from this embedding. This architecture is particularly suited for unsupervised learning, as it can discover compact, low-dimensional representations of high-dimensional data without requiring labeled examples \citep{autoencoders}. In this context, the high-dimensional input corresponds to the 13 chemical abundances measured for each star. However, due to the underlying physics of stellar nucleosynthesis, many of these abundances are correlated, so the data effectively lies on a lower-dimensional structure, which is captured by the latent space. Moreover, autoencoders can serve a denoising purpose: if noise in the input data cannot be captured as a consistent pattern in the latent space, the decoder’s reconstruction effectively produces a cleaner version of the input. Classically, the latent space of an autoencoder is interpreted as capturing the most informative features of the dataset, while filtering out redundancies and noise.\par
More concretely, we implemented a GAT autoencoder with two encoding (or "hidden") layers containing 50 neurons each that map the input node features into a five-dimensional latent space. The choice of five dimensions was motivated by the need to reflect the different nucleosynthesis channels present in the data --- $\alpha$, iron-peak, $s$-process, $r$-process --- plus an additional free dimension, and we empirically found that 50 neurons per layer provided sufficient expressive power to achieve good final reconstruction. The network uses \texttt{GATv2} \citep{gatv2} layers with learnable attention coefficients to dynamically weight neighboring nodes. During training, a \texttt{NeighborLoader} from \texttt{PyTorch Geometric} \citep{pythorch} samples mini-batches of nodes along with their three-hop neighborhoods (15, 10, 5 neighbors per layer), enabling efficient training on large graphs. The decoder mirrors the encoder structure and reconstructs only the node features (i.e., the chemical abundance space) from the latent embeddings. This design ensures that the network learns to assign attention weights to edges in a way that optimizes the reconstruction, effectively identifying which neighboring nodes are most informative for each star. A visual representation of the full algorithm can be seen in Figure~\ref{fig:GAT}. \par

We trained the model for 200 epochs, dividing the graph into five batches, with an initial learning rate of \texttt{lr = 0.008}. The learning rate was dynamically reduced during training using a \texttt{ReduceLROnPlateau} scheduler, which halved the learning rate if the reconstruction loss did not improve for 20 epochs, with a minimum allowed value of 0.0005. The model typically reached a final reconstruction loss of around 0.16 (see Figure~\ref{fig:loss}), where the reconstruction loss was computed as the mean squared error between the input and reconstructed node features. While training the network for 200 epochs, embeddings for analysis were selected based on a calibration graph of known stellar structures (Table \ref{tab:substructures}), choosing the epoch that minimized inter-cluster edges while preserving intra-cluster edges (see Section \ref{subsec:calibration}). For this optimal epoch, we extracted the latent and reconstructed spaces and repeated the process ten times, performing ensemble clustering with the OPTICS algorithm to assign stars membership probabilities for each stable cluster (see section \ref{subsec:ensemble}). This membership reflects both how many times each individual star was associated with a cluster, and how many times a cluster was recovered across all ten experiments. For a deeper understanding of these training and clustering procedures, we refer readers to Appendix~\ref{sec:analysis}.

\section{Data}
\subsection{Observational data}
\label{sec:observational}

To chemically disentangle the various substructures present in the halo, it is crucial to have elemental abundances covering as many nucleosynthesis channels as possible. Our analysis relies on GALAH DR4 \citep{galah}, which, based on high-resolution spectroscopy ($R \sim 28{,}000$), provides detailed chemical abundances for up to 32 elements across nearly one million stars in the MW. A major improvement over DR3 is the new data reduction and analysis pipeline, which implements a hybrid spectrum-synthesis approach supported by NNs trained on synthetic spectra. This method allows stellar parameters and chemical abundances to be determined simultaneously across the full GALAH wavelength range. The synthetic grids used to train the NNs account for non-local thermodynamic equilibrium (NLTE) line formation for 14 elements, including H, C, N, O, Na, Mg, Al, Si, K, Ca, Mn, Fe, and Ba. To complement abundances and stellar parameters, GALAH DR4 also provides a value-added catalog cross-matched with Gaia DR3, which includes kinematic and dynamical information. \par
\begin{figure}[h!]
\centering
\includegraphics[width=\columnwidth]{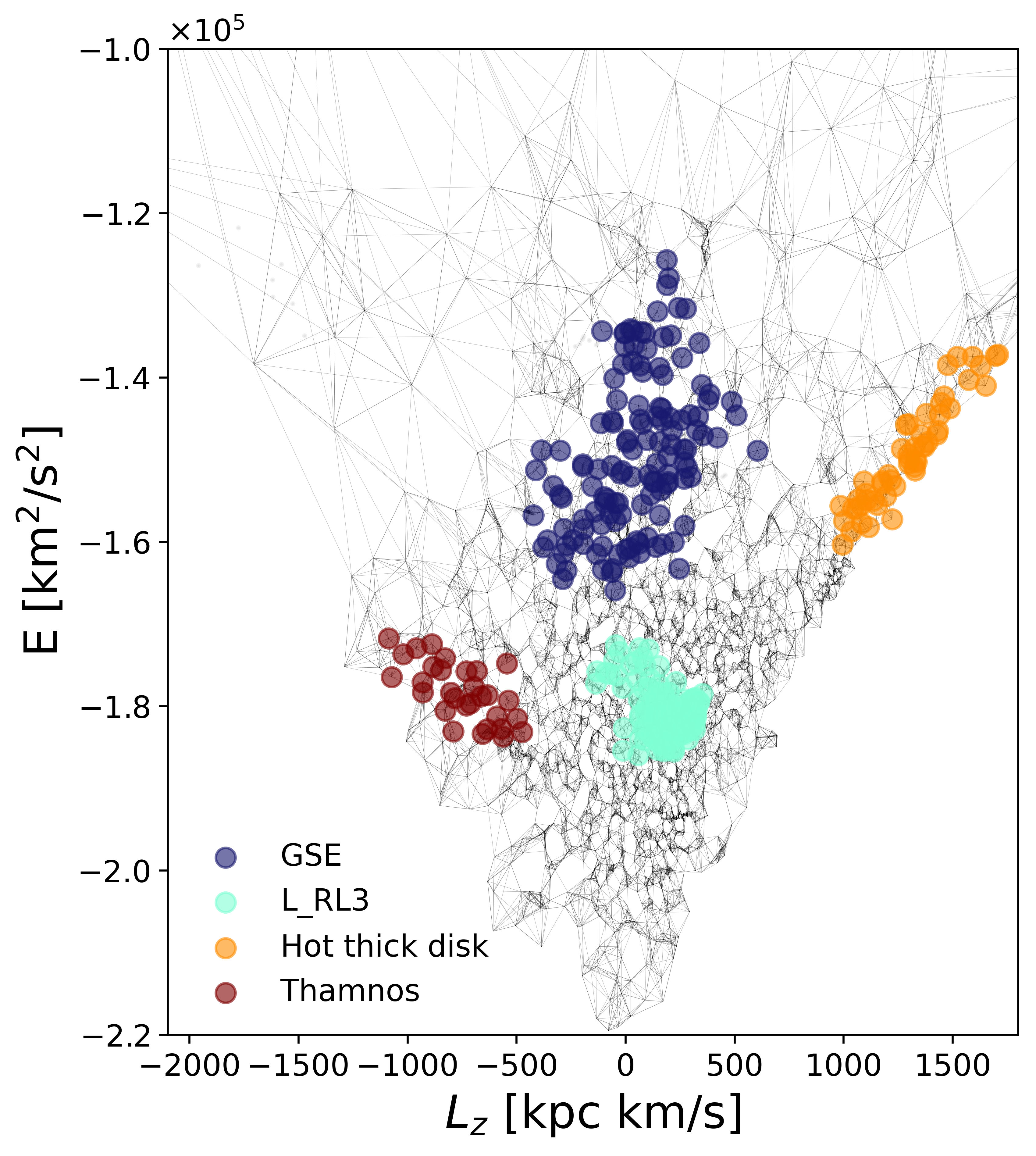} % fits the width of the column
\caption{Lindblad diagram of the halo star sample used in this work. The four target streams identified in \citetalias{Dodd_2023} are highlighted in color, while GCs are shown as gray points for clarity, indistinguishable from field stars. Edges of the graph are represented as gray lines, in which each edge links two stars from the selected GALAH DR4 sample}.
\label{fig:graph}
\end{figure}
To select halo stars while maximizing the number of labeled substructures, we adopted the halo selection of \citet{Dodd_2023} (hereafter \citetalias{Dodd_2023}), yielding 8,887 stars. This selection included stars with $|\mathbf{v}-\mathbf{v}_{\rm LSR}| > 210\,\mathrm{km/s}$ and $d \le 5\,\mathrm{kpc}$ (obtained by inverting parallax), together with standard Gaia quality cuts on parallax precision (relative uncertainty $<20\%$),  \texttt{RUWE}$<1.4$, and line-of-sight velocity uncertainty $\sigma(V{_{\rm los}})<20\,\mathrm{km/s}$. In their case, this selection yields a total of 193,831 Gaia DR3 stars. By applying a single-linkage hierarchical clustering algorithm to discover statistically significant overdensities (regions with enhanced stellar density compared to mock datasets with shuffled velocities) in the space of $E$, $L_z$, and $L_{\perp}$, they identify seven main dynamical groups by retaining structures that exceed the background by 3$\sigma$. Among these, GSE is the most populous, with 7,618 stars, while L-RL3, the Hot Thick Disk, and Thamnos contain 5,005, 2,556, and 1,550 stars, respectively. 174,488 are left unclustered. We then applied the recommended quality cuts by the GALAH collaboration, namely stars with S/R \texttt{snr\_px\_ccd3 > 30}, stellar parameters for stars with \texttt{flag\_sp == 0} and retained only elements reliably measured, that is, with \texttt{flag\_X\_fe == 0}, in at least 55\% of this halo sample, resulting in 13 abundances: Fe, Ca, Mg, Ti, Sc, V, Mn, Cu, Zn, Ba, Y, and Nd. This threshold was set such that we could include critical elements like Ba and Nd. The resulting sample comprises 3,598 stars, including 357 cataloged as streams.
To validate and calibrate our method, we further enriched our sample with globular clusters (GCs). We used the catalog from the Survey "Stellar Clusters in 4MOST" (Lucatello et al. in prep.) and included only those clusters for which our sample contained at least eight members. After all these selections and quality cuts, all stars in the final halo star sample have a good quality measurement for each of the 13 selected chemical abundances. The final sample contains 3,885 stars and its breakdown can be seen in table \ref{tab:substructures}.

\begin{table}[h!]
\centering
\begin{tabular}{lclclc}
\hline
\multicolumn{3}{c}{GCs} & \multicolumn{3}{c}{Streams} \\
\hline
Name & & Count & Name & & Count \\
\hline
NGC6656 & & 108 & Field & & 3105 \\
NGC6121 & & 98  & GSE   & & 132 \\
NGC5139 & & 82  & L-RL3  & & 122 \\
NGC5904  & & 17  & Hot Thick Disk   & & 52 \\
NGC6809 & & 16  & Thamnos   & & 30 \\
\hline
NGC288  & & 19  & Sequoia  & & 7 \\
NGC362  & & 28  & S\_10   & & 5 \\
NGC2808 & & 27  & S\_5   & & 5 \\
NGC6544 & &  20   & S\_6    & & 4 \\
NGC1851 & & 8   &       & &    \\

\hline
\end{tabular}
\caption{GCs and stellar streams represented in the dataset. The horizontal line divides the target (top) and calibration (bottom) sets. Streams labeled as S\_X correspond to streams labeled as X in \citepalias{Dodd_2023}. Field stars are halo stars that appear in neither \citetalias{Dodd_2023} nor the cluster catalog.}
\label{tab:substructures}
\end{table}

The horizontal line in table \ref{tab:substructures} refers to a further distinction done in our analysis: the most numerous groups form the "target set", groups hidden in our dataset that will serve to assess both the performance of our method and also the comparison with the approach of employed in \citetalias{Dodd_2023}. The second group comprises the "calibration set", needed to overcome a methodological aspect that we introduce in the appendix \ref{subsec:calibration}.

\subsection{Graph structure}

After obtaining a clean dataset with all the relevant chemo-dynamical information, we organized the data into a graph. To build the graph edges, we scaled the $E$ and $L_z$ values using a \texttt{StandardScaler} and connected each star to its ten nearest neighbors in this space using the Euclidean distance metric. The choice of ten neighbors ensured that even in regions of the IoM space where data was sparse, particularly at higher energies, each star remained adequately connected. Furthermore, the edges are undirected: if star \textit{i} is linked to star \textit{j}, star \textit{j} is also linked to star \textit{i}. The final graph contains 3,885 nodes and 69,320 edges, shown in Figure~\ref{fig:graph} \footnote{Note that this graph contains both target and calibration sets}. This graph was trained with the GAT as introduced in the previous section to obtain a dynamics-informed, reconstructed abundance space, which was then clustered with OPTICS \citep{optics}.

\begin{figure}[ht]
    \centering
    \begin{subfigure}[t]{\columnwidth}
        \centering
        \includegraphics[width=\textwidth]{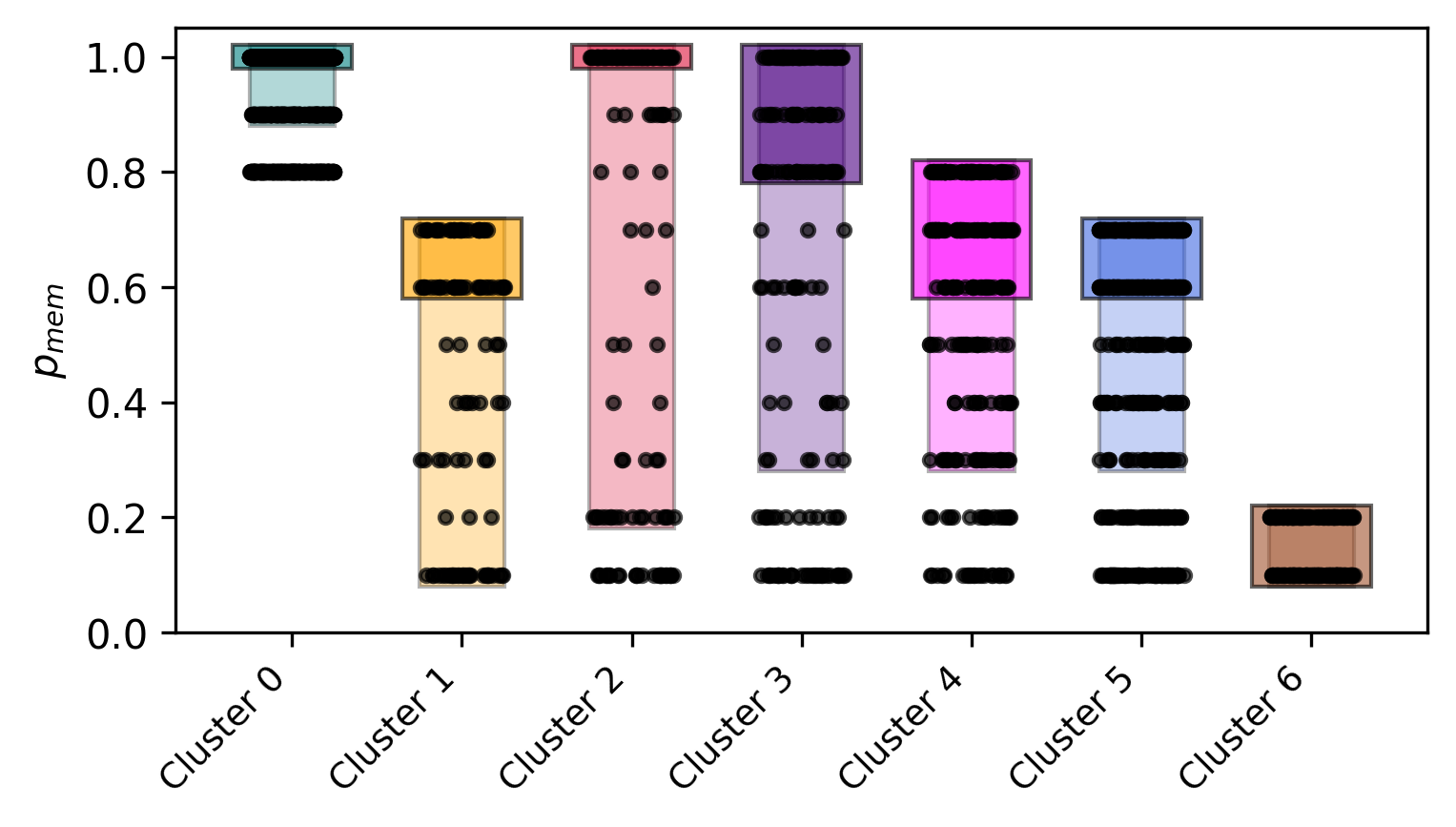}
    \end{subfigure}

    %\vspace{0.5cm}

    \begin{subfigure}[t]{\columnwidth}
        \centering
        \includegraphics[width=\textwidth]{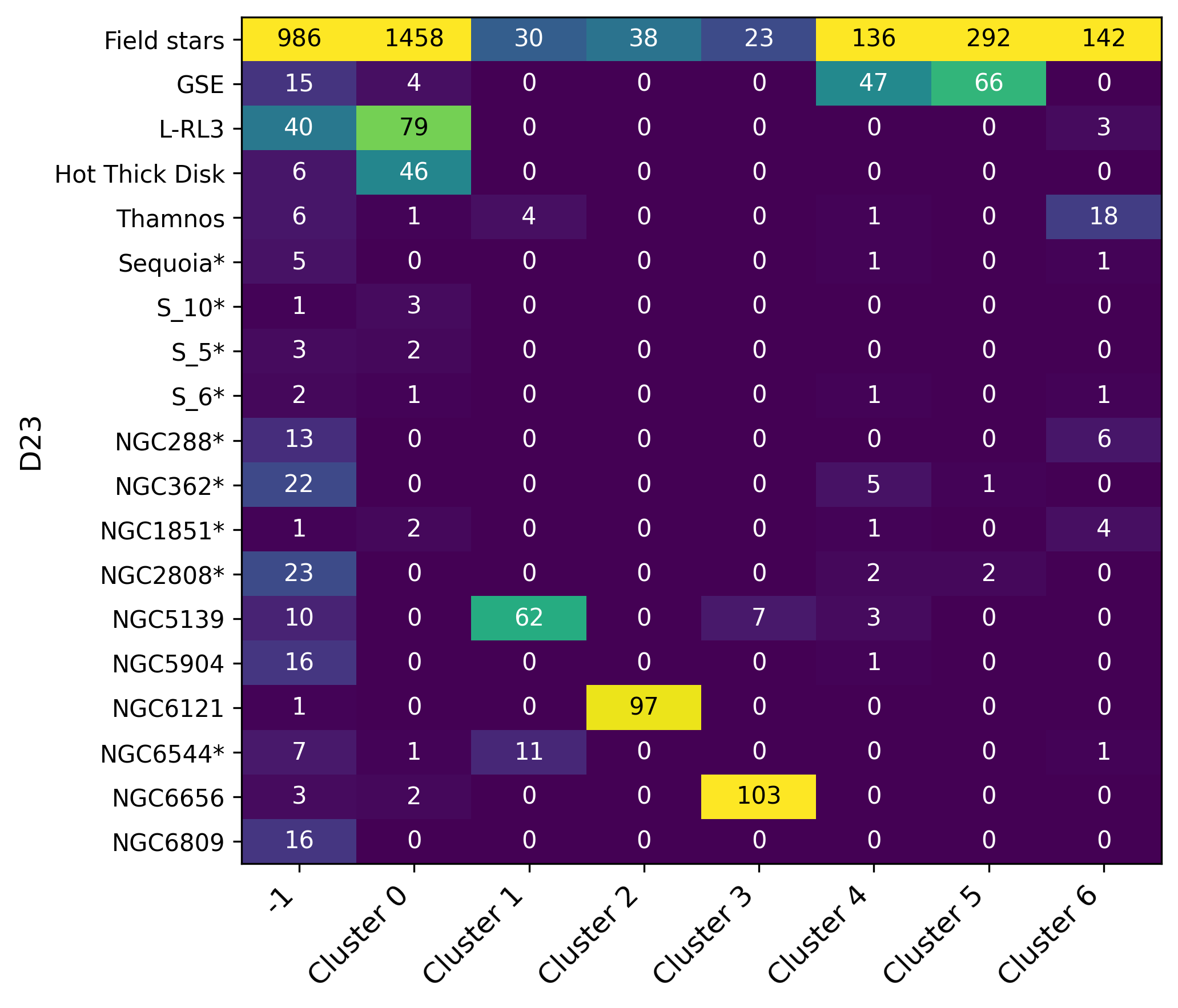}
    \end{subfigure}

    \caption{Top: Membership probability distribution for the different clusters. For each cluster, the light rectangle shows the top 75\% memberships, and the darker rectangle spans the top 50\% (median). Individual stars are overplotted as black points. Bottom: Cross-tab of cluster labels derived in this work versus those from \citetalias{Dodd_2023} and Lucatello et al. in prep.. The heatmap shows the number of stars in each label combination.}
    \label{fig:calib}
\end{figure}

\begin{figure*}[h!]
    \centering
    \includegraphics[width=\textwidth]{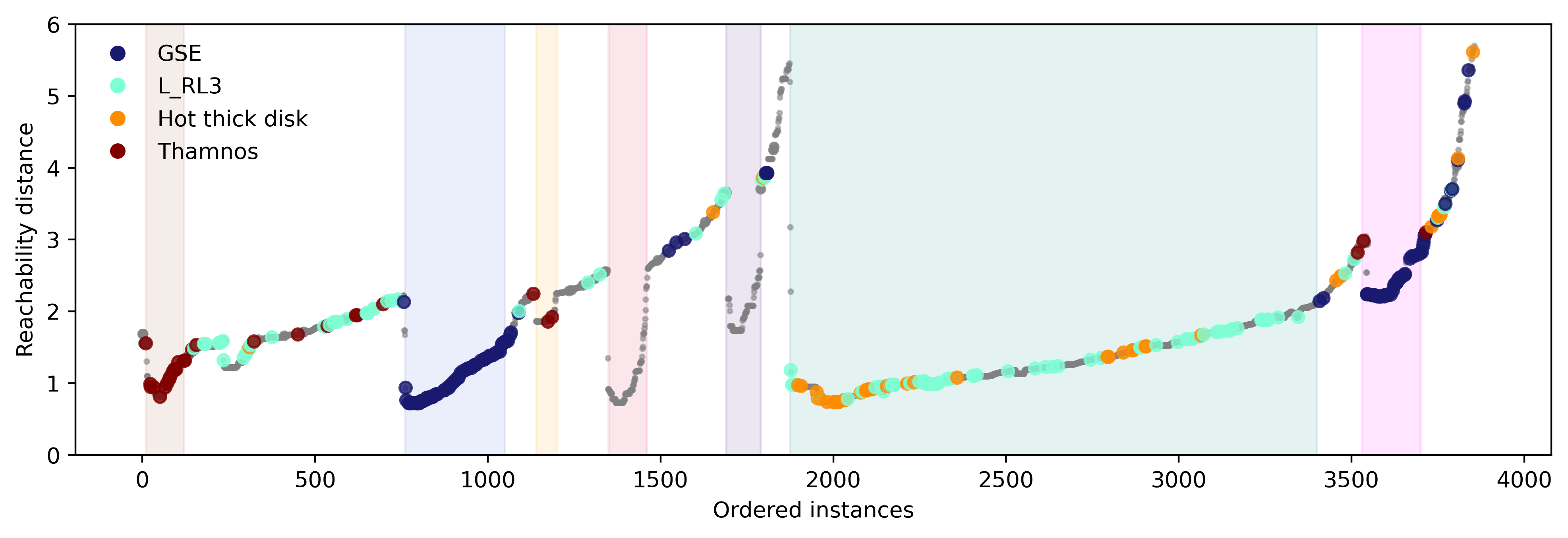}
    \caption{Reachability plot of all identified clusters. Streams from \citetalias{Dodd_2023} are colored according to the palette used in their work. Gray points represent field stars, among which GCs are hidden. Shaded areas indicate the approximate locations of the consensus clusters identified in this work.}

    \label{fig:optics}
\end{figure*}

\begin{figure*}[h!]
    \centering
    \includegraphics[width=\textwidth]{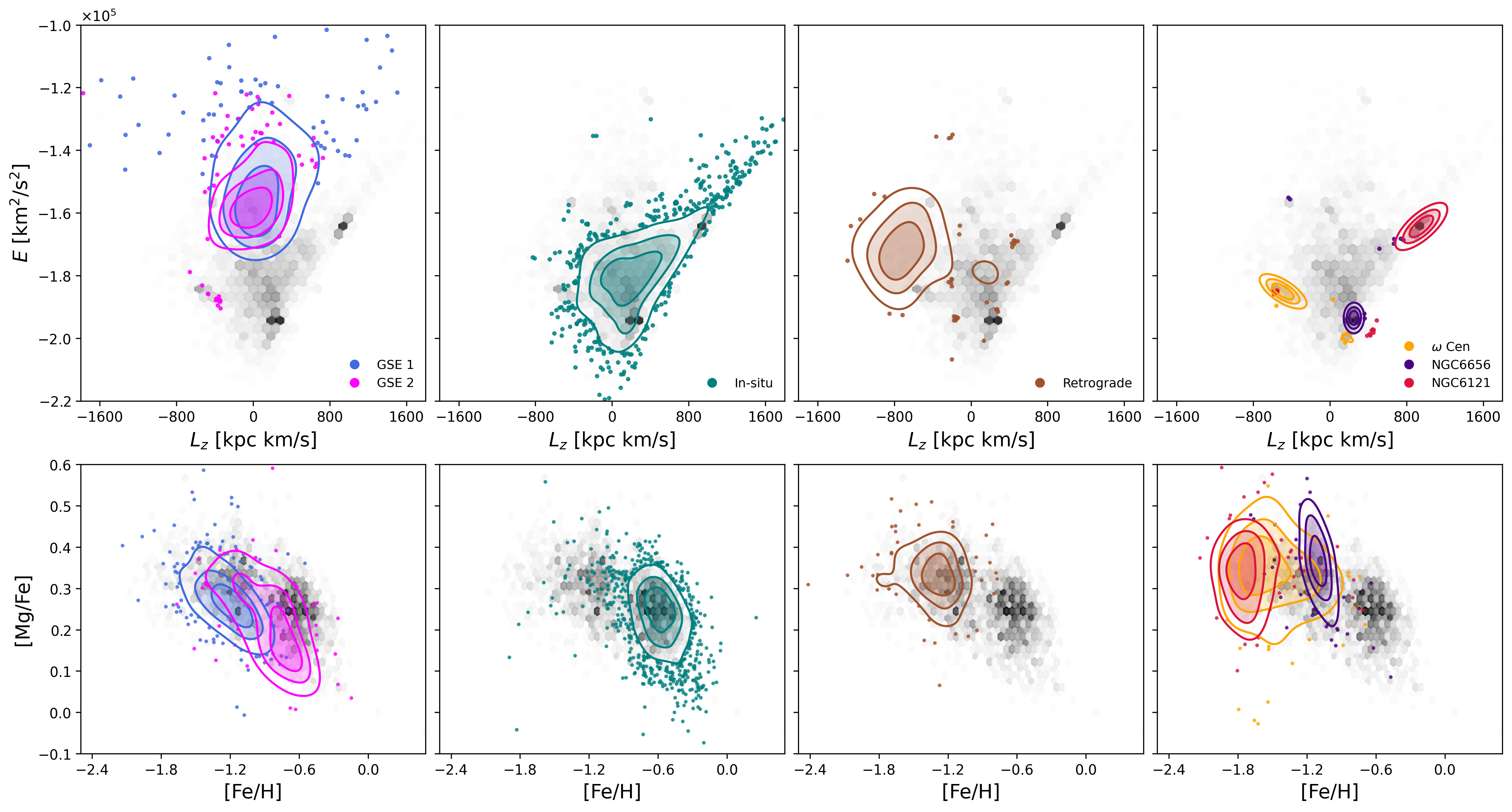}
    \caption{Top: Lindblad diagram for all clusters identified in this work. All stars are shown in gray (hexbin), while clusters are overplotted using a 2D Gaussian KDE. The three highest-density (60, 80 and 100\%) levels are shown with filled contours, and stars outside these regions are plotted as scatter points to highlight lower-density members and outliers. Bottom: [Mg/Fe] vs [Fe/H] abundance plane for the same clusters, using same criteria for the density contours. }

    \label{fig:clusters}
\end{figure*}

\section{Results and discussion}
\label{sect:results}

In this section, we discuss the seven clusters identified in the GAT reconstructed abundance space. All results correspond to the best training epoch, selected using the calibration procedure described in Appendix~\ref{subsec:calibration}, and include only clusters deemed reproducible by the ensemble clustering pipeline in Appendix~\ref{subsec:ensemble}. We consider the top 75\% membership probabilities for each cluster. To assess the recovery of GCs, which provide the only available ground-truth labels in our sample, we quantify (i) the homogeneity $H$ and (ii) the completeness $C$ of each recovered cluster. For a recovered cluster $K$ and a globular cluster $GC$, these are defined as $H = |K \cap GC| / |K|$ and $C = |K \cap GC| / |GC|$, respectively. The cluster breakdown and membership probabilities are shown in Figure~\ref{fig:calib}, together with the OPTICS reachability plot in Figure~\ref{fig:optics}. In this plot, each identified cluster corresponds to a distinct dip in reachability distance, with its location highlighted by a shaded colored region.

\subsection{Retrieved substructure}

\begin{figure*}[h!]
\centering
\includegraphics[width=\textwidth]{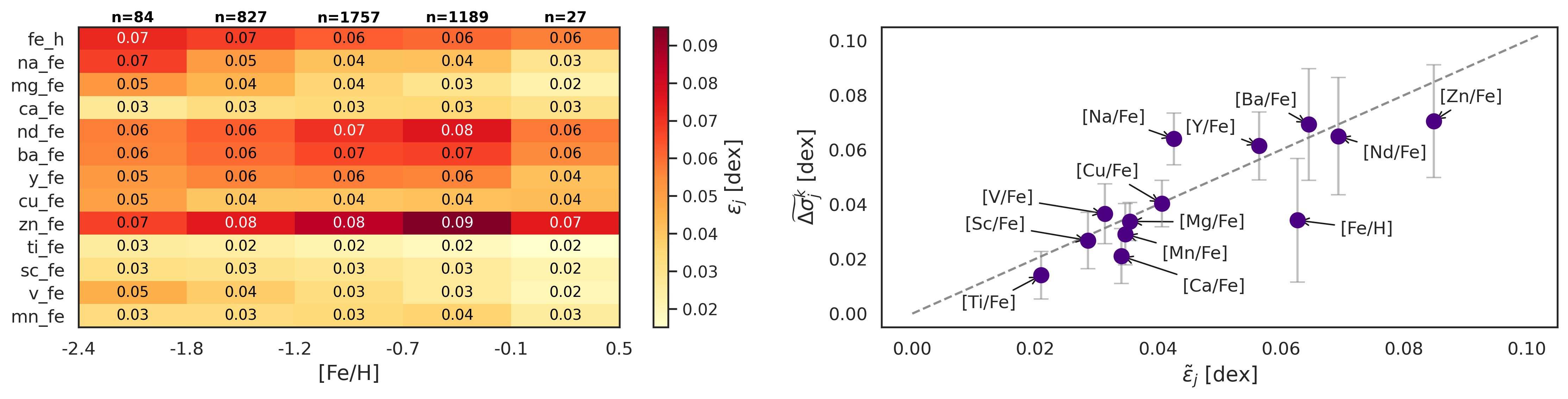}
\caption{Left: Median observational errors per element and metallicity bin. On top of each column, we show the number of stars that are in each bin in our dataset. Right: Median reduction in scatter across ten perturbed reconstruction spaces versus observational errors. Gray dashed line follows the 1:1 relation, and the error bars represent the standard deviation.  }
\label{fig:denoising}
\end{figure*}

\begin{itemize}

    \item Cluster 0: This cluster corresponds to the group extracted in the first step of the clustering pipeline and is attributed to the in-situ (or \textit{splash}) component of the halo for several reasons. First, its relatively high metallicity combined with high [$\alpha$/Fe] ratios is consistent with expectations for an in-situ halo population, aligning with previous studies \citep{Matteo2018TheMW, nissen2023abundancesironpeakelementsaccreted, mori2025chemicaldecodingkinematicsubstructures}. It is also by far the most numerous group, with 1,599 stars (see top Figure~\ref{fig:calib}) comprising approximately 41\% of the dataset—a fraction consistent with both observational constraints \citep{Bonaca_2017, Naidu_2020} and simulations \citep{cooper_2015, Khoperskov_2023_1}. Most of the Hot Thick Disk stream (88\%) falls within this cluster, as expected. Another relevant structure is L-RL3, first reported by \citet{Lovdal_2022}, a prograde, low-energy group with a median metallicity of [Fe/H] $\simeq -0.70$ \citep{Bellazini_2023}. Given its high [Fe/H], both \citet{Ruiz_Lara_2022} and \citet{thomas2025} suggest that this stream is primarily composed of in-situ stars, with some contribution from accreted systems. This interpretation is consistent with our results: 64\% of L-RL3 stars are classified as in-situ, while the remaining stars are unclustered. No other known stellar streams contribute a significant fraction of stars to this cluster. From now on, we refer to this cluster as 'in-situ'.

    \item Cluster 1 contains 107 stars, of which 62 correspond to NGC5139, or simply $\omega$Cen, leading to a recovery of this cluster with a $H$ and $C$ of 0.58 and 0.76, respectively. Additionally, there are three stars from the Thamnos stream that are consistently clustered with this group (with probabilities of 0.6, 0.6, and 0.3, respectively). Interestingly, \citet{mori2025chemicaldecodingkinematicsubstructures} already speculate on a possible connection between these two systems. In their analysis, they find that the chemical outliers of Thamnos display peculiar abundance trends, in particular a prominent branch at high [Al/Fe], which is atypical for accreted dwarf galaxies but characteristic of GCs. Although Thamnos is sparsely represented in our dataset and [Al/Fe] was not used in our analysis, our clustering results still support this potential connection. On the other hand, in a very recent paper \citet{ceccarelli2025walkretrogradewrsproject} have attributed this high $\alpha$ and Al abundances to high in-situ contamination present in the stream. We attribute this consensus cluster to $\omega$Cen.

    \item Cluster 2 contains 135 stars, of which 97 correspond to NGC6121, while the rest are field stars, yielding a recovery of this GC with $H = 0.72$ and $C = 0.99$. We therefore attribute this cluster to NGC6121. This cluster is highly reproducible, passing the Jaccard threshold in all ten runs. 
    
    \item Cluster 3 contains 103 stars from NGC6656, yielding an excellent recovery with $H = 0.77$ and $C = 0.95$. This cluster also includes seven stars from $\omega$Cen, and 23 field stars for a total of 133. In all ten runs, this cluster matched at least five other runs at the minimum Jaccard threshold. We therefore attribute this cluster to NGC6656.
    
    \item Cluster 4 contains 198 stars, of which 47 (36\%) are associated with the GSE according to \citetalias{Dodd_2023}. The majority, 136 stars, are field stars, and the remaining 15 belong to various mixed substructures. This cluster also appears to be highly reproducible, passing the Jaccard threshold in eight out of ten runs. From now on, we refer to this cluster as 'GSE 2'.
    \item Cluster 5 contains 361 stars, including a significant fraction of \citetalias{Dodd_2023} GSE members (66 stars). Most of the cluster, 292 stars, are field stars, with only 3 belonging to GCs. This cluster is recovered in seven out of ten runs. From now on, we refer to this cluster as 'GSE 1'.
    
    \item Cluster 6 contains the bulk of the retrograde halo, including most of the stars of Thamnos (16 stars). Figure~ \ref{fig:calib}, however, shows a maximum membership probability of 0.2 for this cluster, indicating that this consensus cluster was built from only two stable clusters. In other words, only two runs analyzed as in Section \ref{subsec:ensemble} produced clusters with at least five matches at the minimum Jaccard threshold of 0.5. This result suggests that the cluster is less robust and more diffuse than the others, probably due to the sparsity of the retrograde halo region. For this reason, we do not draw further conclusions about this cluster in the present work. We refer to this cluster as 'Retrograde'.
\end{itemize}

We successfully recover the three most numerous GCs (NGC6656, NGC6121, and NGC5139) with excellent $H$ and $C$ parameters. In contrast, NGC5904 and NGC6809 remain entirely unclustered, as shown in the cross table in Figure~\ref{fig:calib}. This outcome reflects our choice of hyperparameters; recall from Table \ref{tab:optics_params} in Step 2 that the \texttt{min\_cluster} parameter is set to 30, that is, at least 30 stars are needed to form a cluster. This value was specifically chosen to avoid detecting smaller, spurious groups, since our primary interest is in studying the larger structures. \par

In Figure~\ref{fig:clusters}, we show the Lindblad diagram and [Mg/Fe]–[Fe/H] distributions for the different clusters, organized according to the halo substructure they most closely trace: GSE, in-situ, the retrograde halo, and GCs. The in-situ cluster predominantly occupies the prograde, low-energy region of the Lindblad diagram, with a distribution very similar to the one found by \citet{buder_2025} (their Figure~5 top).  We also note that the overlap between the GSE-related clusters and the in-situ cluster is mainly found around $E = [-1.6, -1.8]\cdot 10^5$ km$^2$/s$^2$, while the presence of GSE stars at lower energy values is essentially absent (except for a slightly retrograde clump near the location of $\omega$Cen). More importantly, the number of stars chemically compatible with the GSE system appears to span a much wider range of IoM than the ones identified using the algorithm of \citetalias{Dodd_2023} (see also Figure~ \ref{fig:donlon}). This behavior, previously noted by \citet{berni} and \citet{mori2025chemicaldecodingkinematicsubstructures}, suggests that chemical abundances can reveal potential GSE members across the IoM plane, even if they have lost their dynamical coherence. The majority of these stars were previously classified as field stars in \citetalias{Dodd_2023}, meaning they were not associated with any dynamical substructure or GC. Our chemical tagging method identifies them as chemically compatible with GSE, indicating that GSE 1 and GSE 2 include both the previously identified GSE stars plus additional candidates that have lost dynamical coherence but retain the characteristic abundance patterns.

\subsection{Denoising}
\label{subsec:denoising}
\begin{figure}[h!]
\centering
\includegraphics[width=\columnwidth]{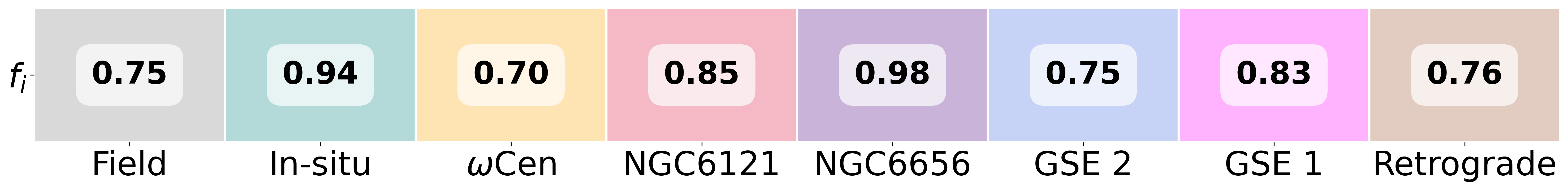} % fits the width of the column
\caption{The per-cluster stability ratio between the labels retrieved from running the experiment with and without perturbing the input abundances.}
\label{fig:clusters_denoising}
\end{figure}

Before drawing further conclusions, we assess the robustness of these results with respect to observational errors and quantify the reduction in spread between the original and reconstructed abundances. We achieve this by training ten independent models, each receiving a slightly perturbed version of the input chemical abundances. The perturbations for each star $i$ were drawn from a Gaussian distribution with standard deviations equal to the reported observational errors for each element $j$, i.e., $x_{i,j}^{\rm pert} = x_{i,j}^{\rm orig} + \mathcal{N}\big(0, \epsilon_{i,j}^2\big)$, where $\epsilon_{i,j}$ is the measurement error given by GALAH DR4. We show the distribution of these errors in metallicity bins in the first panel of Figure~\ref{fig:denoising}. We applied the ensemble clustering pipeline and recovered the same seven stable clusters as in the previous subsection. The \emph{stability fraction} of each cluster $K_i$ is defined as the fraction of stars that remain in the same cluster after perturbing the input abundances, $f_i = |K_i \cap \hat{K}_i|/{|K_i|}$, where $\hat{K}_i$ denotes the corresponding cluster in the perturbed clustering: a value of one would indicate identical clusterings. The per-cluster $f_i$ is shown in Figure~ \ref{fig:clusters_denoising}, demonstrating that not only the same clusters are found, but also that the result is in very good agreement in all cases.  \par

Furthermore, we assess whether the reduction in scatter of each reconstructed abundance is consistent with its observational uncertainties. We define the reduction in scatter as: 
\begin{equation}
    \Delta \sigma_{j}^{k} = \sigma_{j}^{\rm obs} - \sigma_{j}^{\rm pert, k}  
\end{equation}

where $\sigma_{j}^{\rm obs}$ is the standard deviation of element $j$ in the original observed abundances, and $\sigma_{j}^{\rm pert, k}$ is the standard deviation of the reconstructed abundances for the $k$-th perturbed input. We took the median, $\widetilde{\Delta \sigma}_{j}^{k}$, across all perturbed reconstructions and compared it to the median observational error, $\tilde{\epsilon}_j$. The results shown in the right panel of Figure~\ref{fig:denoising} indicate that the reconstruction is stable and generally reduces noise to a level comparable with and for most elements slightly smaller than observational errors. The main exception is [Na/Fe], for which the reduction in scatter is approximately 1.5 times larger than its typical observational uncertainty. This behavior is likely caused by the correlations of [Na/Fe] with [V/Fe] ($r=0.46$) and [Cu/Fe] ($r=0.39$). When these correlated abundances define a consistent trend, deviations in [Na/Fe] are more likely to reflect measurement noise than intrinsic variation. As a result, the reconstruction pulls [Na/Fe] toward the trend defined by these elements, reducing its scatter more noticeably than elements that do not have strong, focused correlations. The error bars denote the standard deviation of this reduction across reconstructed spaces, which range from 0.007 dex in [Mg/Fe] to 0.023 dex in [Fe/H].\par
Overall, the fact that the same consensus clusters are also recovered under perturbations, and that the reduction in scatter correlates with the typical abundance errors, suggests that the GAT is not overfitting the noise, but rather effectively denoising the chemical abundance space. \par

\subsection{The GSE dichotomy}
\label{subsec:gse_dichotomy}

In the previous subsections, we outlined the general characteristics of the clusters, connected them to several known or proposed halo substructures and accounted for their variability under perturbations. For the rest of the paper, we focus on the observed dichotomy of the potential GSE stars. We emphasize that the emergence of these two clusters is fully data-driven, since the model does not have any information about the labels. As such, GSE 1 and GSE 2 naturally reflect groups of stars that share similar IoM but have systematically different chemical abundances. In this subsection, we explore the origin of this dichotomy from the perspective of stellar nucleosynthesis, and test whether it reflects a real astrophysical signal rather than an artifact of the reconstruction. \par

\begin{figure}[h!]
\centering
\includegraphics[width=\columnwidth]{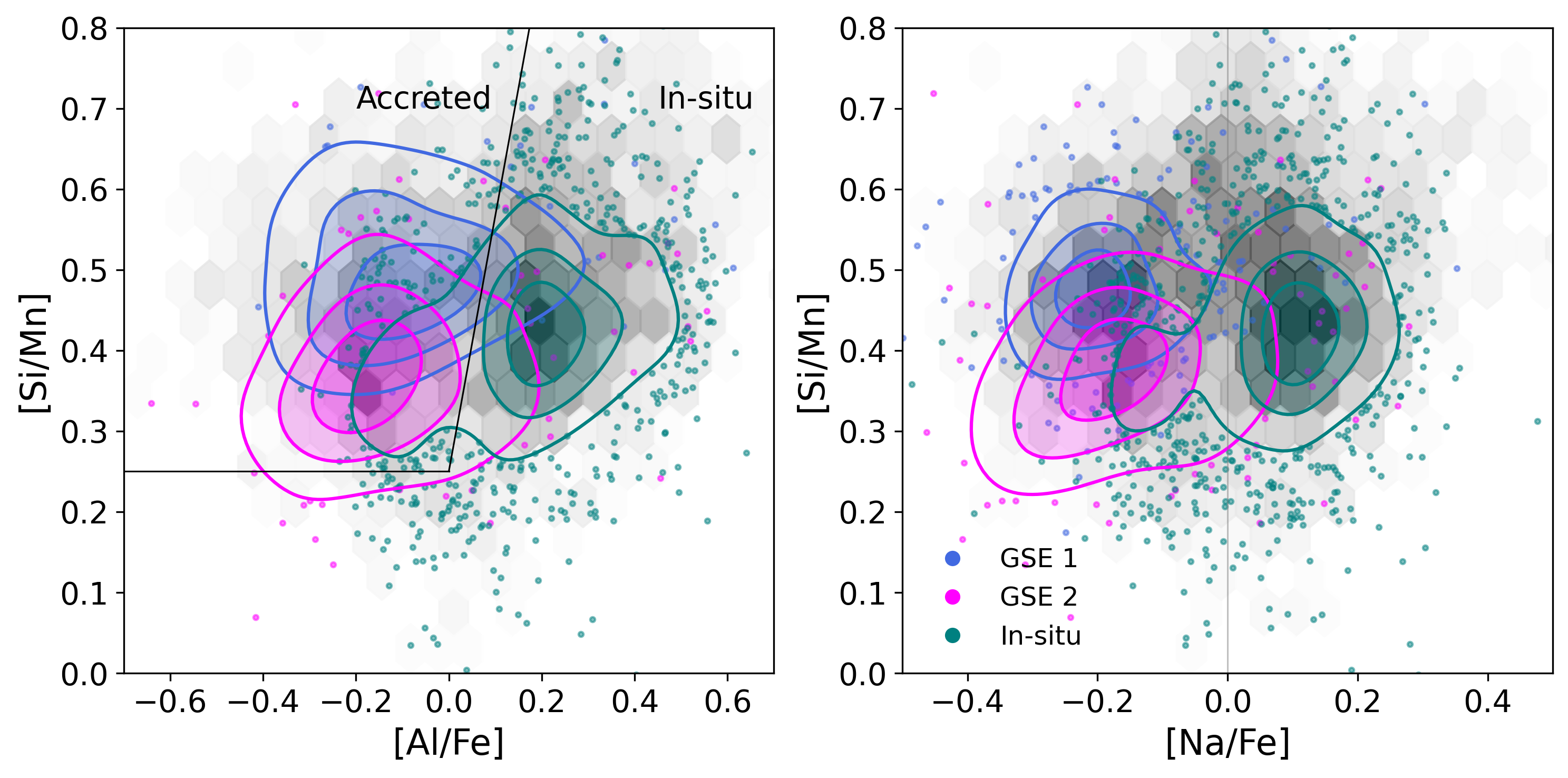} % fits the width of the column
\caption{Left:[Si/Mn] vs [Al/Fe] abundance plane for the two clusters associated with GSE and the in-situ sample. The three highest-density levels are shown with filled contours (spanning from 60\% to 100\% of the peak KDE density), while stars outside these regions are plotted as scatter points to highlight lower-density members. Black lines denote limits commonly adopted in the literature. Right: Same for [Si/Mn] vs [Na/Fe], with a gray line at solar.}
\label{fig:aluminium}
\end{figure}

We begin by examining the distributions of GSE 1 and GSE 2 relative to the in-situ clusters in the [Si/Mn]–[Al/Fe] plane (Figure~\ref{fig:aluminium}). This diagnostic space is particularly effective for distinguishing accreted from in-situ populations, as first demonstrated by \citet{hawkins_2015}, and later used in many other studies \citep{das, horta_2021, fernandes_2022}. However, given the known challenges in measuring Al from optical spectra in GALAH, especially for metal-poor stars \citep{heiter_2019}, and the fact that previous GALAH-based studies have often favored Na as a more robust alternative \citep{buder_2021}, we also show an analogous diagram using [Na/Fe] in place of [Al/Fe].

Importantly, we did not use Al abundances in the GAT training, making this an independent test of our methodology. To further validate our approach with another previously unused element, we adopt Si abundances instead of the more commonly used Mg, but still a canonical $\alpha$ element. All stars shown in Figure~\ref{fig:aluminium} were selected according to the quality cuts described in Section \ref{sec:observational}, leaving 96 stars for GSE 1, 141 stars for GSE 2, and 1544 for the in-situ population. Measuring Al in metal-poor stars is notoriously challenging, as previously noted by \citet{Roederer_2021}, and is the main cause of the sharp decline in reliable measurements for our clusters, particularly for GSE 1. Despite the limited number statistics, both clusters associated with GSE occupy the region characteristic of accreted, chemically unevolved systems, while the in-situ cluster lies predominantly in the evolved region (although a branch extends into the unevolved region). This branch is slightly reduced when using Na to improve statistics, and the compactness of GSE 1 is similarly enhanced. This reinforces the idea that our methodology can successfully identify accreted stars even without relying on Al abundances, and that both GSE 1 and 2 are of accreted origin. \par
This naturally raises several key questions: Which elements are primarily responsible for driving this separation between GSE 1 and GSE 2? If such a distinction exists, can we still attribute both clusters to the same accreted system? \par

\begin{figure*}[p]  % 'p' = float on a dedicated page
    \centering
    \includegraphics[width=1.0\textwidth]{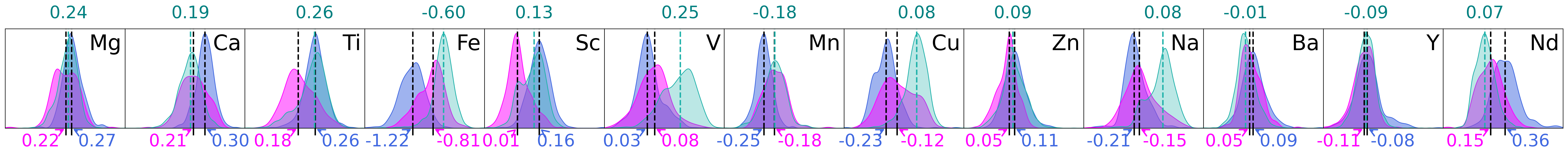}
    \includegraphics[width=0.9\textwidth]{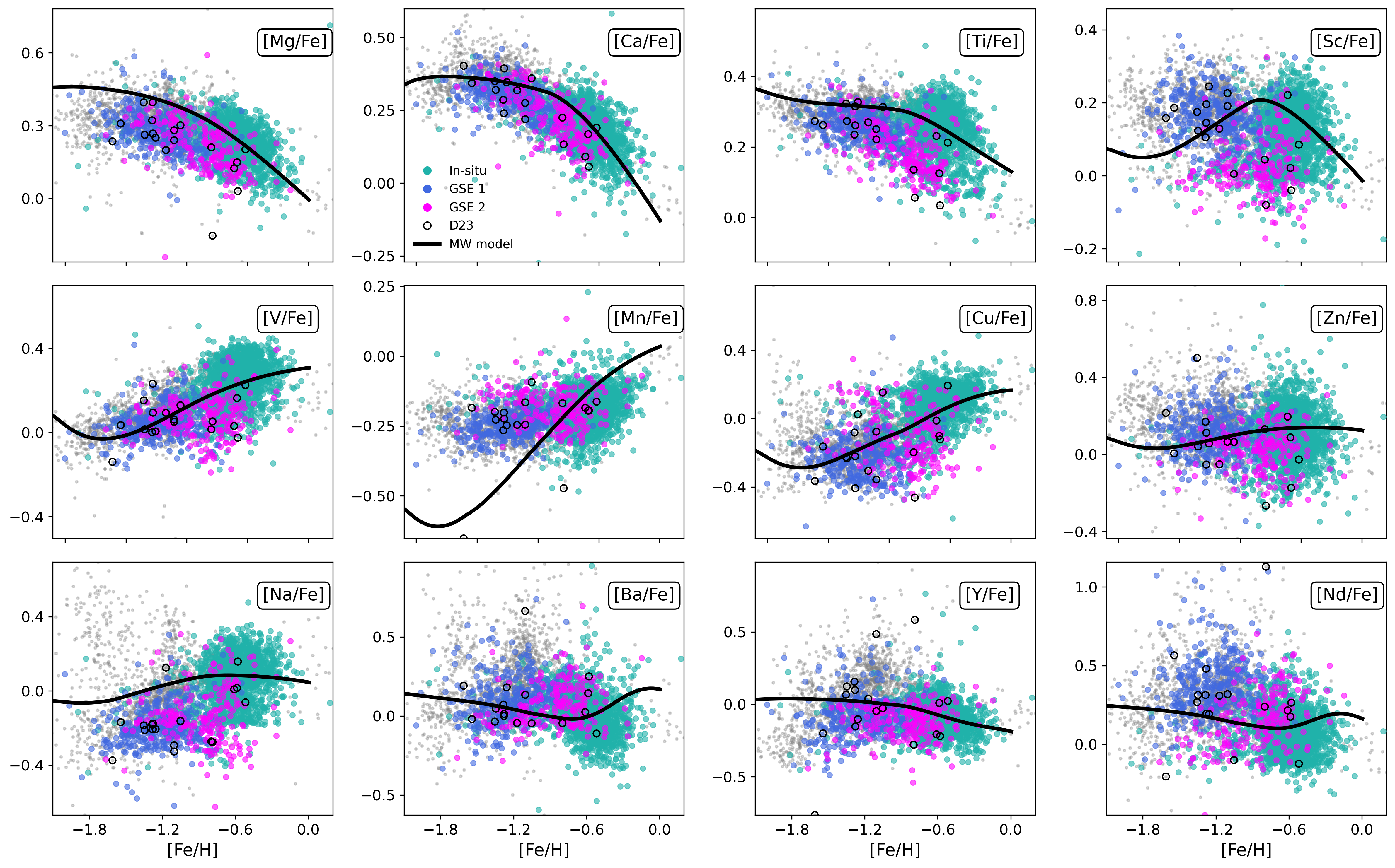}
    \includegraphics[width=0.9\textwidth]{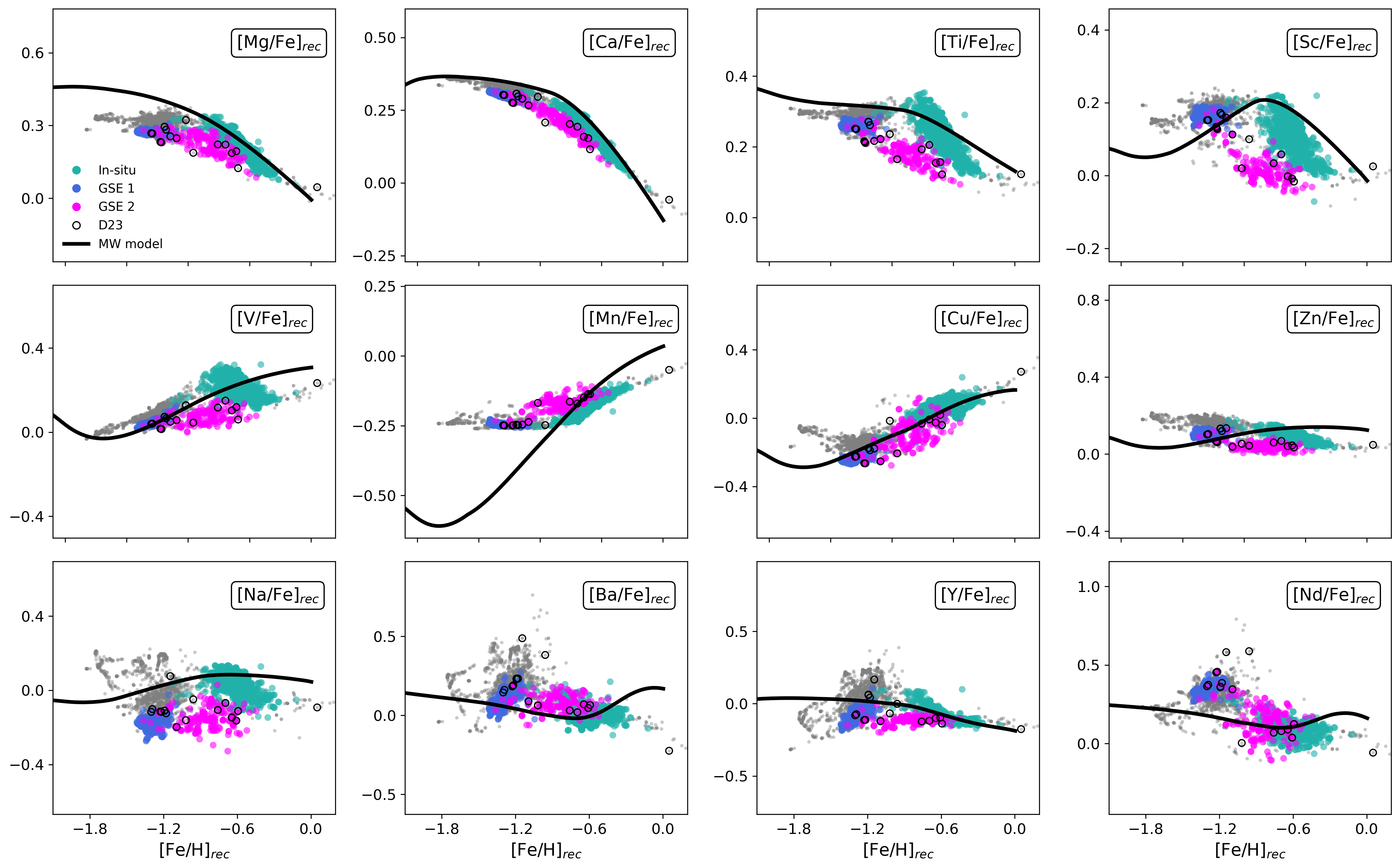}
    \caption{ Top: KDE histograms and median abundances (in [dex], dashed lines) for all elements across the GSE and in-situ clusters. Middle: Original elemental abundances compared to Galactic chemical evolution models \citep{romano_2010, romano_2019}; models are $y$-shifted to match the Hot Thick Disk baseline. Background stars are shown in gray, while colored points indicate cluster members. Black-edged symbols represent stars identified as GSE in \citetalias{Dodd_2023} but not recovered in this work. Bottom: Same as middle panel, but utilizing GAT-reconstructed abundances.}
    \label{fig:all_elements_vs_model}
\end{figure*}

To better understand this and identify which elements drive the separation, we show all elements used in this work in Figure~\ref{fig:all_elements_vs_model}, presenting both the original measurements and the GAT-reconstructed values. The 19 stars labeled as GSE members in \citetalias{Dodd_2023} but missing from our analysis are shown as empty circles, which could be either contaminants present in \citetalias{Dodd_2023} (\citet{mori2025chemicaldecodingkinematicsubstructures} for instance estimate the contamination of some dynamical GSE definitions at 22\%), stars with outlier abundances in some of the chemical dimensions, or stars with lower membership probability than the one employed to analyze the results. 
We also provide a comparison with a MW chemical evolution model, which is a revised version of the model for the MW halo and discs discussed in \citet{romano_2010, romano_2019}. The stellar yields for low- and intermediate-stars are taken from the FRUITY database \footnote{\href{http://fruity.oa-teramo.inaf.it/phys_modelli.pl}{FRUITY database}}, while those for massive stars are taken from the ORFEO database\footnote{\href{http://orfeo.iaps.inaf.it/}{ORFEO database}}. We refer the reader to papers by \citet{cristallo_2015, Cristallo_2016, limongi} and \citet{roberti}  for a description of the stellar evolutionary models and to \citet{spitoni} and Romano et al. (in prep.) for details about, respectively, the Galactic chemical evolution (GCE) model and the yield implementation. The model assumes a two-infall scenario, but here we show only the first 3.35 Gyr, corresponding to the first infall phase. We also caution the reader that the model has been shifted in the $y$-axis to match the median abundances of the Hot Thick Disk, using a weighted least-squares fit between the model and the observed medians in bins of [Fe/H]. We summarize in what follows the (original) chemical trends observed for GSE 1 and GSE 2, and compare them to the in-situ. For a visual comparison, we show the distributions and median values of each cluster in the top panel of Figure~\ref{fig:all_elements_vs_model}. 
\begin{itemize}
    \item \textbf{$\alpha$ elements}: Both clusters form a coherent decreasing sequence in the Mg, Ca, and Ti versus [Fe/H] plane, with abundances systematically lower than the MW at fixed metallicity, consistent with a less efficient star formation history. Notably, GSE 1 is more $\alpha$-enhanced than GSE 2. 

    \item \textbf{Iron-peak elements}: V and Mn exhibit an almost flat, slightly rising trend with [Fe/H], while Zn is slightly enhanced in GSE 1 relative to GSE 2. This Zn enhancement is consistent with its production in hypernovae at low metallicities \citep{kobayashi}. We also note that Mn abundances are higher than the MW predictions for most GSE stars, while they align much better with the flatter Zn and V predictions.

    \item \textbf{Odd-Z elements}: The Na and Cu depletion is apparent for both GSE components when compared to the in-situ sample. This was already noted in previous works, such as \citet{nissen_schuster_2011, carrillo}. While Cu abundances in GSE generally follow the predictions, Na falls clearly below. Sc, on the other hand, closely follows the decreasing trend of Ti and visually appears to be one of the main drivers of the separation between GSE 1 and GSE 2. This $\alpha$-like behavior of Sc has been reported in previous studies for both MW stars \citep{battistini_2015} and satellite systems such as Sextans \citep{theler_2020}. 

    \item \textbf{Neutron-capture elements}: Y exhibits an almost flat trend with [Fe/H] and is largely indistinguishable from the in-situ population. Ba, on the other hand, shows a slightly decreasing trend, with values slightly higher than those of the MW. Nd, in contrast, is enhanced in GSE 1 compared to GSE 2 and, like Sc, visually appears to be a major driver of the separation between the two components.
\end{itemize}

\begin{figure*}[h!]
\centering
\includegraphics[width=\textwidth]{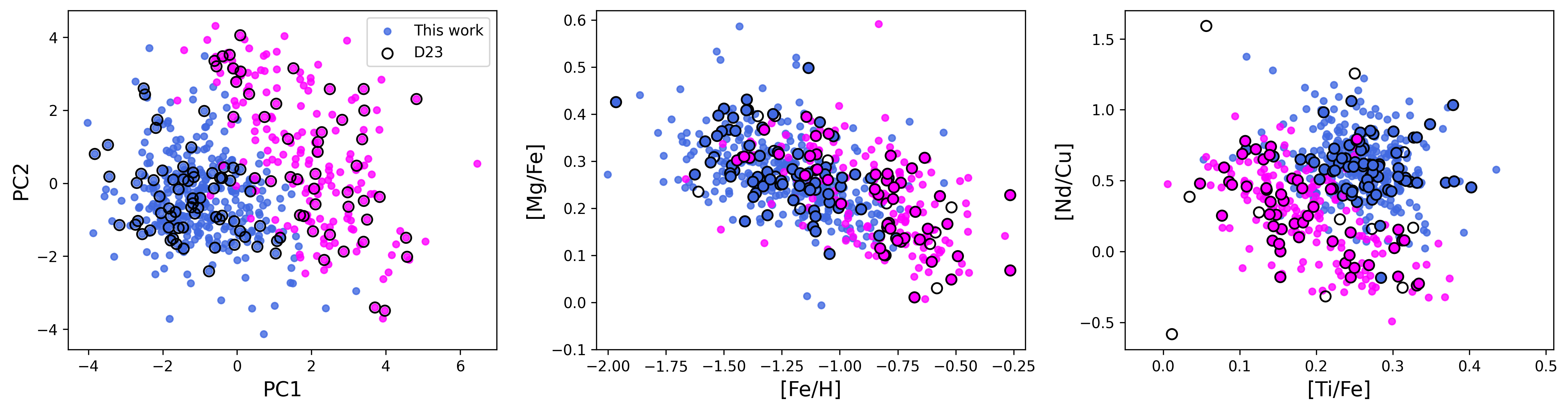}
\caption{Left: PCA projection onto first two components. The stars from this work are shown in color and GSE members from \citetalias{Dodd_2023} are highlighted with black edges.
Middle: [Mg/Fe] vs [Fe/H] plane.  \citetalias{Dodd_2023} stars appear highlighted with black edges, now also showing the missing stars from their work. 
Right: Same, but for the [Nd/Cu] vs [Ti/Fe] plane. }
\label{fig:pca}
\end{figure*}

The reconstructed abundances obtained with the GAT reveal several notable features relative to the original measurements (Figure~\ref{fig:all_elements_vs_model} bottom). First, some reconstructed abundances closely follow the trends predicted by the MW chemical evolution model, indicating that the autoencoder captures the underlying chemical evolution patterns. This is particularly evident for Ca, Mg, Ba, Y, Cu, and Mn. Second, the reconstructed abundance planes are consistently narrower than the original measurements, suggesting that the GAT effectively reduces scatter and suppresses observational uncertainties while preserving physical trends, as has been already quantified in subsection \ref{subsec:denoising}. A clear example is [Mg/Fe], which not only tracks the model closely but also exhibits a significantly narrower distribution. The same behavior is observed for the two GSE clusters, which follow similar trends but at lower [Mg/Fe] values. Finally, the GAT autoencoder enhances chemical distinctions between stellar populations. Visually, for example, the original [Sc/Fe] and [Ti/Fe] abundances already exhibit strong discriminating power between populations. The autoencoder then captures these differences and further enhances them, making the separation between populations even more pronounced in the reconstructed space. \par
Because the autoencoder operates in an unsupervised manner, it identifies patterns based solely on the structure present in the input data. However, the algorithm only aims at reducing the  loss. While this enables it to integrate information from the IoM and to denoise and emphasize dominant trends, it remains essential to verify that these patterns are also present---albeit more subtly---in the original abundance space. Otherwise, there is a risk of over-interpreting features emphasized by the model’s training objective rather than by the underlying stellar chemistry. \par

To independently verify that the chemical dichotomy between GSE 1 and 2, is present in the original abundance data, we perform a principal component analysis (PCA) (see section \ref{appendix_3} for a further test using a GMM). PCA is an unsupervised dimensionality-reduction technique that identifies the linear combinations of features along which the data variance is maximized. If subpopulations of GSE stars differ systematically in multiple elemental abundances, PCA will align the first few components with these directions of greatest variance, allowing the high-dimensional separation between subpopulations to become visible in a low-dimensional projection. The result of the PCA applied to the original space, showing the first two principal components of GSE 1 and GSE 2, is presented in Figure~\ref{fig:pca}. The separation between the two groups is evident and aligns very well with the two clusters detected by the GAT. To assess which elements contribute most to these components, we present the PCA loadings in Table \ref{tab:loadings}. These loadings quantify the relative influence of each element on the variance captured by the components, highlighting which abundances are primarily responsible for the separation between the GSE groups.

\begin{table}[h!]
\centering
\begin{tabular}{lrr}
\hline
Element & PC1 & PC2 \\
\hline
\textnormal{[Fe/H]}   &  0.454 &  0.093 \\
\textnormal{[Mg/Fe]}  & -0.350 &  0.248 \\
\textnormal{[Ca/Fe]}  & -0.416 &  0.109 \\
\textnormal{[Ti/Fe]}  & -0.446 &  0.199 \\
\textnormal{[Sc/Fe]}  & -0.388 & -0.114 \\
\textnormal{[V/Fe]}   &  0.001 &  0.431 \\
\textnormal{[Mn/Fe]}  &  0.241 &  0.332 \\
\textnormal{[Cu/Fe]}  &  0.075 &  0.513 \\
\textnormal{[Zn/Fe]}  & -0.240 &  0.078 \\
\textnormal{[Na/Fe]}  &  0.018 &  0.370 \\
\textnormal{[Ba/Fe]}  &  0.026 & -0.169 \\
\textnormal{[Y/Fe]}   & -0.105 &  0.117 \\
\textnormal{[Nd/Fe]}  & -0.128 & -0.348 \\
\hline
\hline
Explained variance & 0.27 & 0.18 \\
\hline
\end{tabular}
\caption{PCA loadings for the GSE stars on the first two principal components.}
\label{tab:loadings}
\end{table}

PC1, which accounts for the largest fraction of the variance (27\%) and along which the two GSE groups are primarily separated, is dominated by $\alpha$ elements such as [Mg/Fe], [Ca/Fe],  [Ti/Fe], and [Sc/Fe], all with strong negative loadings, as well as [Fe/H] and [Mn/Fe] with positive loadings.

PC2 captures a smaller fraction of the variance (18\%) and is dominated by [Cu/Fe], [V/Fe], and [Na/Fe] with positive loadings, and by [Nd/Fe] and [Ba/Fe] with negative loadings, suggesting that this axis primarily traces variations in neutron-capture and odd-Z elements. The metal-poor component, GSE 1, forms a compact concentration in PCA space, while the more metal-rich GSE 2 appears more chemically extended. This distinction is mildly visible in the classical [$\alpha$/Fe]–[Fe/H] planes, where both groups follow the typical declining sequence expected for dwarf galaxy chemical evolution (middle panel). A much stronger separation emerges in the [Nd/Cu]–[Ti/Fe] plane (rightmost panel), where GSE 1 occupies higher [Ti/Fe] and [Nd/Cu] values and forms a tight, coherent cluster, whereas GSE 2 shows lower ratios and a broader dispersion. This behavior can be understood in terms of the nucleosynthetic origins of these elements. Ti and Sc primarily trace enrichment from core-collapse (CC) supernovae and are therefore sensitive to the early star formation history of the progenitor. Nd is mainly produced by the $r$-process at $\mathrm{[Fe/H]} \lesssim -1$, reflecting stochastic enrichment in low-mass systems, with an increasing contribution from the $s$-process at higher metallicities as AGB stars begin to contribute. By contrast, Na and Cu are odd-$Z$ elements with strongly metallicity-dependent yields: Na is produced in massive stars with increasing yields at higher metallicity, while Cu has a complex origin involving both massive stars and Type~Ia supernovae and shows a marked decline at low metallicity \citep{Kobayashi_2022}. These distinct nucleosynthetic sensitivities imply that the observed abundance gradients within the GSE system are tracers of its star formation efficiency: The  [$\alpha$/Fe], [Sc/Fe], and [Nd/Fe] enhancement in GSE 1 are consistent with the products of early, less chemically evolved and CC-dominated enrichment, while the more metal-rich GSE 2 reflects a more efficient and sustained star forming regime with increased contributions from Type~Ia supernovae and AGB stars.
\par 
It is worth noting that the Cu I transitions most commonly used to derive copper abundances are significantly affected by NLTE effects (see, e.g., \citealt{korotin_2018}; \citealt{yan_2016}), particularly at low metallicities. However, in the metallicity regime of our sample, NLTE corrections are modest \citep{caliskan_2025} and cannot fully account for the observed differences between populations, suggesting the depletion found in this work is real and consistent with previous works \citep{nissen_schuster_2011, carrillo}.

\begin{figure*}[h!]
\centering
\includegraphics[width=\textwidth]{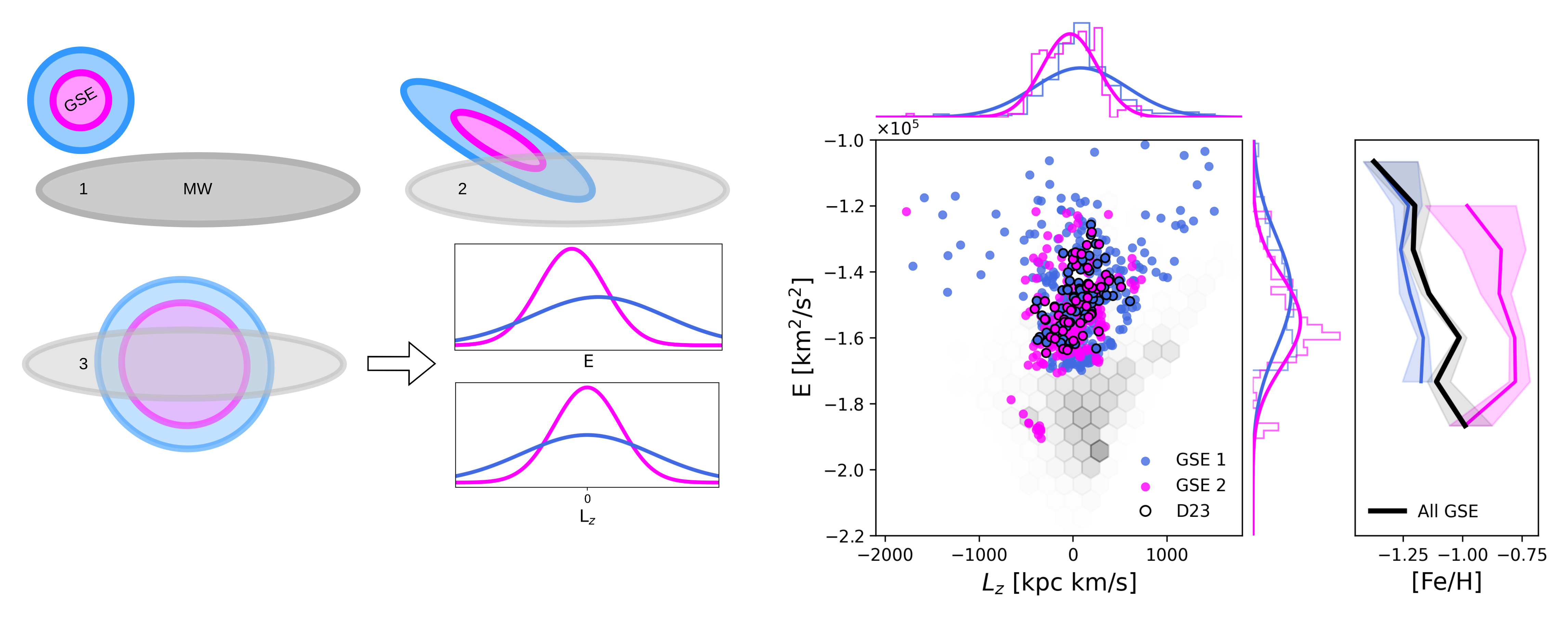}
\caption{Left: Co-infall scenario proposed by \citet{donlon_2023} for an ancient massive accretion event, together with the qualitative $E$ and $L_z$ distributions such an event would produce.
Right: (1) $E$ and $L_z$ distributions of GSE 1 and GSE 2 determined in this work (black edges mark the GSE selection in \citepalias{Dodd_2023}). The histograms show normalized probability density distributions with Gaussian fits overlaid. (2) The bootstrapped [Fe/H] median trends (over 1000 samplings) for GSE 1, GSE 2, and the combined GSE 1+2. Shaded areas indicate the 16th–84th percentile uncertainties from the bootstrap. }
\label{fig:donlon}
\end{figure*}

\subsection{Comparison to previous studies}
\label{sec:comparison}

The presence of a dichotomy in the GSE IoM region was first hinted at by \citet{berni}, who identified two discrete clumps using a GCN autoencoder trained on [M/H] and [$\alpha$/Fe], which appear vertically stacked in the Lindblad diagram. The differences with respect to the present work may reflect a methodological distinction: the fixed topology of a GCN constrains message passing to the initial IoM-based edges, whereas the learnable edges in the GAT allow the network to adjust attention weights based on chemical similarity. \par

\citet{carrillo} identify GSE candidate stars by fitting a GMM to the [Mg/Mn]–[Al/Fe] plane and selecting the component that lies in the unevolved, accreted region of this diagram. For these stars, they measure high-resolution ($ 22{,}000$–$40{,}000$) abundances for 20 elements, including iron-peak, $\alpha$, and both $s$- and $r$-process neutron-capture elements. Following the terminology introduced by \citet{das}, they refer to this component as the \textit{blob}. By comparing its detailed abundance patterns with the low-$\alpha$ halo sample of \citet{nissenshuster}, they find general chemical agreement between both samples. However, \citet{carrillo_2024} show that stars selected in the [Mg/Mn]–[Al/Fe] plane can occupy different regions in dynamical space, indicating that the chemical selection may also include stars with diverse origins. The blob stars exhibit lower [$\alpha$/Fe] than MW disk stars at similar metallicities ([Fe/H] $\gtrsim -1.5$~dex), along with depleted Na, Ni, and Cu, indicative of slower chemical evolution. In contrast, neutron-capture elements (e.g., Ba, La, Ce, Nd, Y, Zr, Eu) show $r$-process enhancement and delayed $s$-process enrichment.  We note how GSE 1 (peaking at [Fe/H] $\gtrsim -1.2$~dex, slightly higher than the blob at [Fe/H] $\gtrsim -1.6$~dex), follows the general trends of the blob stars, while the higher-metallicity component, GSE 2, aligns better with the low-$\alpha$ NS \citep{nissenshuster} sample, at least qualitatively. 
\par

\citet{Sk_lad_ttir_2025} propose a two-passage scenario for the GSE merger event, in which stars from the outer regions of the progenitor, accreted during an early pericentric passage, end up at higher $E$ in the Lindblad diagram, while stars from the inner regions, accreted during a later passage, occupy lower $E$ orbits. 
In parallel, while not necessarily implying a two-passage scenario, \citet{buder_2025} show—using NIHAO-UHD simulations \citep{buck_2018} —that the present-day chemodynamical and spatial properties of accreted stars do in fact retain a memory of their birthplaces in the progenitor: chemically evolved stars formed in the core occupy lower energies, whereas less evolved stars from the outskirts are stripped first and thus reside at higher energies. The most notable difference between their work and ours is that we do find metal-rich, potential GSE members at relatively high energies, whereas their study claims that such stars are predominantly confined to much lower energies, in the inner regions of the MW.

\subsection{A possible infall scenario}

We have discussed how the two GSE clusters are consistent with a less chemically evolved and a more chemically enriched one. To round up, we draw inspiration from \citet{donlon_2023} (see their sections 4.1 and 4.2) and present a very simplified model that can explain the observed chemodynamical trends, connecting both clusters found in this work to their birthplaces within the progenitor. \par
In a merger scenario where the outer regions of the progenitor remain bound until collision with the MW, both low- and high-[Fe/H] stars initially share similar $E$ and $L_z$ momentum and energy. However, the low-[Fe/H] stars are located in the progenitor's outer disk (in our work, GSE 1), have a larger internal velocity dispersion and occupy a broader range of positions. As a result, after the merger, these stars exhibit a wider spread in $E$ and $L_z$ relative to the inner region. Their higher velocity dispersion also translates into somewhat higher energies after accretion (Figure~\ref{fig:donlon}, left). On the other side, the remaining bound core (in our work, GSE 2) continues to lose energy and sink deeper due to dynamical friction, peaking at lower $E$. In reality, as also noted by the authors, it is also expected that the outermost stars of GSE were in part stripped first, perhaps at an early passage, and end up at the highest $E$ values (the scenario proposed by \citet{Sk_lad_ttir_2025}).
The $E$ and $L_z$ distributions retrieved in this work are presented in the right panel of Figure~\ref{fig:donlon} and are in agreement with the scenario proposed by \citet{donlon_2023}. In addition, we show the bootstrapped median [Fe/H] trend with $E$, which increases from $[{\rm Fe/H}] = -1.37^{+0.18}_{-0.04}\,\rm{dex}$ at the highest energies to $[{\rm Fe/H}] = -0.99^{+0.11}_{-0.07}\,\rm{dex}$ at the lowest energies, consistent with a scenario in which the outermost layers of the progenitor are the first to be lost, or in which stars have higher energies due to their larger internal velocity dispersion prior to accretion.

\subsection{Caveats}
We note, however, that these distributions may be biased by selection effects in the dataset. As suggested by \citet{buder_2025}, the most metal-rich component of GSE is concentrated in the inner Galaxy near the bulge. Our sample likely misses many of these stars because GALAH, as an optical survey, has limited coverage in highly extinct regions like the bulge. Moreover, the survey’s magnitude limit restricts our observations to the local halo, so a significant number of outer halo stars belonging to GSE 1 are likely missing. These two limitations would affect the IoM distributions as well as the interpretation of the merger scenario. \par
Furthermore, we have interpreted both components (GSE 1 and GSE 2) as belonging to the same accreted system. Simulations generally indicate that the inner stellar halo is dominated by one or two massive progenitors \citep{purcell_2007, deason_2016}. However, it remains completely possible that these two components correspond to distinct accretion events originating from separate progenitors. \par
Finally, it is clear from the cross-table of in Figure~\ref{fig:clusters} that some contamination is expected for all of our clusters. Clusters that represent the GCs, for instance, have a homogeneity ranging from 0.58 to 0.77. This contamination naturally carries over to the other halo substructures, and while we are not able to quantify it, we caution that some  contamination is indeed expected, especially when including stars with lower memberships. Figure~\ref{fig:aluminium}, on the other hand, shows the in-situ sample partly occupying the unevolved region. This suggests that some high-metallicity accreted stars may be buried at lower energies and remain undetected, or are attributed to the in-situ population. Because these stars are primarily connected to the dominant in-situ population in the graph, their neighbors have conflicting chemical patterns, which highlights a limitation: in regions of the Lindblad diagram where a structure is heavily underrepresented, the method may struggle to reconstruct its properties. We also note that the exact cluster assignments slightly differ between the labeling obtained from the original abundances and that obtained from perturbed inputs. This variation is expected given the intrinsic stochasticity of NNs together with the further variation introduced when clustering each reconstructed space with different hyperparameters and perturbing the input. We account for this variability by providing membership probabilities; consequently, this probabilistic framework reflects the inherent uncertainty in individual star assignments and should guide the interpretation of our results. However, the consistent recovery of the same cluster structure and the clear separability of these groups in PCA abundance projections demonstrate that the method reliably identifies chemically coherent populations, even if individual stellar assignments show some variations.

\section{Conclusions}
In this work, we present a framework that simultaneously exploits chemical abundances and IoM to study the substructures present in the MW halo. To achieve this, we represented the data as a graph, where the nodes correspond to individual stars and their features encode 13 chemical abundances from GALAH DR4 --- Fe, Ca, Mg, Ti, Sc, V, Mn, Cu, Zn, Ba, Y and Nd. By processing the data with a GAT autoencoder, we demonstrated that graph-based architectures enhance the chemical coherence of different stellar groups and, by comparing the reconstructed spaces with GCE models and perturbing the input abundances within observational error range, denoise the original abundances.  \par
We constrain the fraction of in-situ stars to be approximately 41\% in this sample, and find that the Hot Thick Disk and L-RL3 streams are largely part of this population, around 88\% and 64\% respectively. Using our fully agnostic selection, and comparing it to \citet{Dodd_2023}, we find that chemically clustered GSE samples span a wider range of IoM than dynamically clustered ones, comprising up to 15\% of the sample. We find that GSE is depleted in [Cu/Fe], [Na/Fe], and [Al/Fe] compared to the MW, consistent with previous findings \citep[e.g.,][]{nissen_schuster_2011}. Strikingly, these GSE stars further separate into two distinct components: GSE 1 and GSE 2. This separation is driven by clear differences in chemical maturity, with GSE 1 exhibiting significantly higher [$\alpha$/Fe], [Sc/Fe], and [Nd/Fe] ratios. To connect their present-day chemodynamics with their birthplaces in the progenitor galaxy, we examine their $E$ and $L_z$ distributions and propose a simple explanatory model: the outer parts of the progenitor, GSE 1, are less chemically evolved and occupy higher $E$, while the more chemically evolved core GSE 2 is concentrated at lower $E$ with narrower distributions. Finally, we also find that metallicity increases with decreasing $E$, consistent with recent studies suggesting that GSE was a massive disk galaxy with a negative metallicity gradient that dominates the accreted inner halo \citep{Sk_lad_ttir_2025, buder_2025, carrillo_2025}.

In light of upcoming spectroscopic surveys such as WEAVE \citep{weave}, 4MOST \citep{4most}, MOONS \citep{moons}, and DESI \citep{Cooper_2023}, GNN-based methods have proven efficient at extracting information from the chemical space that would otherwise remain inaccessible in the raw data. The approach can be particularly useful for low-resolution spectra, reducing noise and enabling large-scale mining of the chemical abundance space of halo stars \textit{en masse}. The signatures of the Milky Way’s formation history are buried deep in the data, but can be uncovered with the state-of-the-art statistical methods in synergy with astrometric information.

\begin{acknowledgements}
This manuscript has been substantially improved thanks to the constructive comments and suggestions of the anonymous referee. We would also like to thank Donatella Romano for providing the GCE models. This project is funded by the European Union’s Horizon Europe program under the Marie Sk{\l}odowska--Curie Actions (HORIZON-MSCA-2022-DN-01) Doctoral Network EDUCADO, grant agreement No.\ 101119830. Views and opinions expressed are however those of the author(s) only and do not necessarily reflect those of
the European Union. Neither the European Union nor the granting authority can be held
responsible for them.
\end{acknowledgements}

\bibliographystyle{aa}      % A&A BibTeX style
\bibliography{biblio}   % Name of your .bib file (without .bib)

\appendix

\section{Analysis}
\label{sec:analysis}
In this section we address two points: first, a calibration set prevents the model from exploiting self-loops and second, an ensemble clustering pipeline ensures only reproducible clusters are retained.
\subsection{The calibration set}
\label{subsec:calibration}

The analysis, is not conducted in the final reconstructed space. While the inclusion of learnable edge weights is astrophysically motivated, it introduces a shortcut in the training process: because each node includes a self-loop, the network "realizes" it can reduce the loss by just assigning all the attention to itself and none to its neighbors. In that case, the resulting reconstructed space would primarily reflect the original stellar features, and the model would effectively collapse to a standard feed-forward autoencoder, neglecting the relational information encoded in the graph. To prevent this, we employed a calibration set of known stellar structures and implemented early stopping during training. These clusters, with distinct chemical signatures, allow us to monitor intra- and inter-cluster edges and identify the training epoch that best captures genuine chemical similarity among member stars. The overall procedure follows the workflow outlined in Section 4.2 of \citet{spina_2025}. \par

We began by selecting eight stars from each of the calibration GCs listed in Table \ref{tab:substructures}, along with all stars from the four remaining stellar streams. Limiting the selection to eight stars per GC prevents the GCs’ chemical signatures from dominating the resulting edge attention weights. To ensure chemical homogeneity, we restricted the selection to first-generation stars based on their Mg and Na abundances. The remaining stars from the calibration clusters --- i.e., those not identified as first-generation --- were returned to the target set. \par

Next, we constructed the calibration graph by linking all members of a single structure (or known cluster/group) to each other, whose attention weights we call $\alpha_{k}$ (with a total of 488 of these kinematics edges), and additionally linked each star to a randomly selected star from a different calibration structure, with attention weights $\alpha_{r}$ (with a total of 244 random edges). For each node, we then computed the ratio between the attention weight of an intra-cluster edge and its self-loop, $R_k = \alpha_{k} / \alpha_{\text{self}}$, which quantifies how strongly the node attends to other members of its own structure relative to itself. Similarly, the ratio for inter-cluster edges, $R_r = \alpha_{r} / \alpha_{\text{self}}$, measures the relative attention given to stars from other structures. \par

\begin{figure}[h!]
    \centering
    \includegraphics[width=\columnwidth]{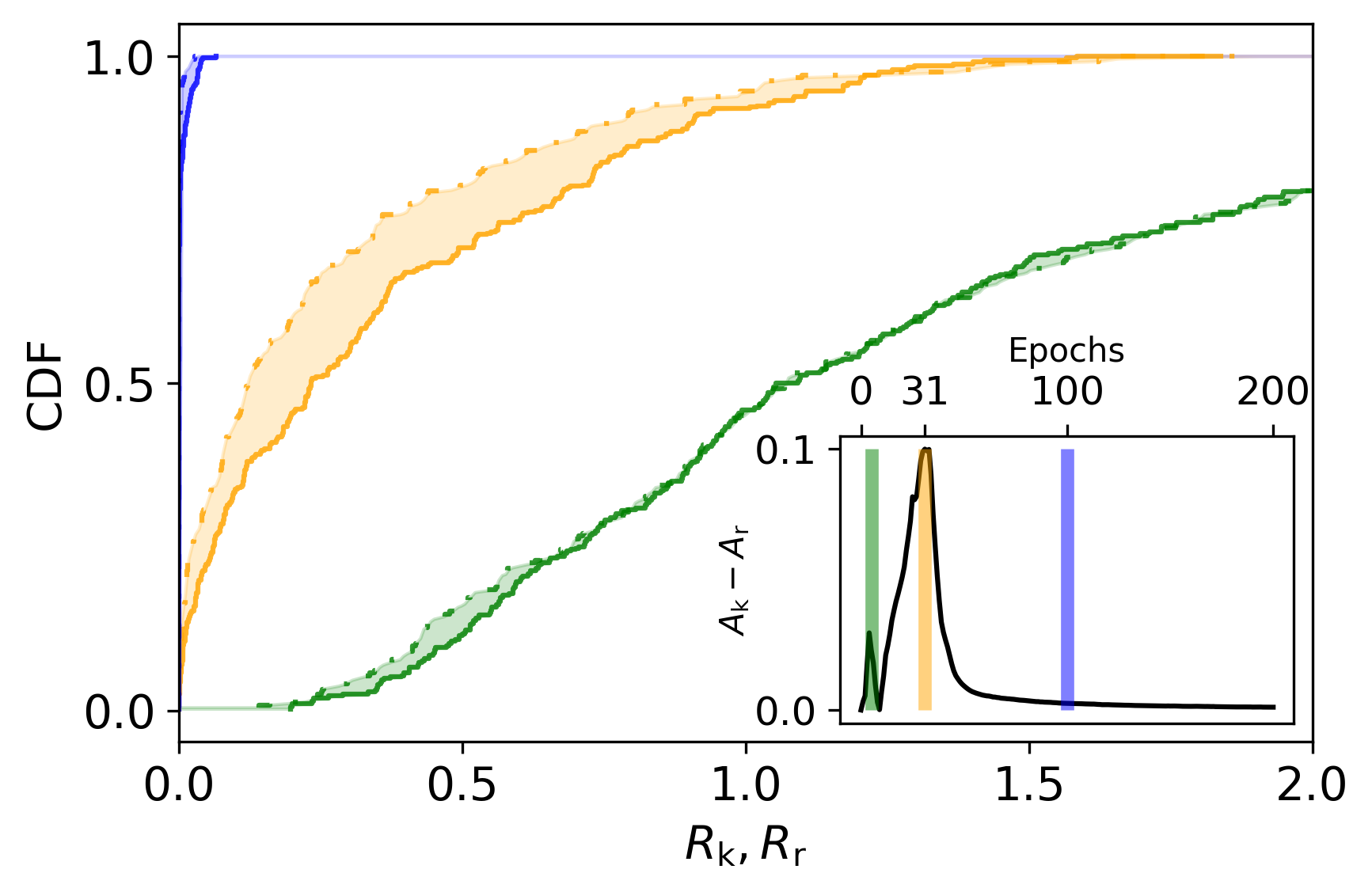}

    \caption{Cumulative distribution functions (CDFs) for $R_k$ (solid line) and $R_r$ (dash-dotted line) at epochs 0 (green), 31 (yellow), and 100 (blue) for one of the ten models analyzed in this work. The inset represents the same epochs, now showing the area between both CDFs. For each run, we select the model that maximizes this area.}
    \label{fig:calib1}
\end{figure}

For our analysis, we focused on models where $R_k$ is consistently larger than $R_r$, ensuring that edge importance reflects chemical similarity while preventing embeddings from being dominated by self-loops. This behavior is illustrated in Figure~ \ref{fig:calib1}, where we plot the per-node cumulative distribution functions (CDFs) of $R_k$ and $R_r$. During the first epoch, both intra-cluster (continuous line) and inter-cluster (dashed line) edges primarily reflect their initial values. As training progresses, edges connecting chemically dissimilar stars are gradually downweighted, while edges between members of the same structure are preserved or reinforced. By epoch 31, the separation between the two edge types reaches a maximum, as quantified by the area between the corresponding CDFs in the inset panel of Figure~\ref{fig:calib1}.

Finally, by epoch 100, both $R_k$ and $R_r$ decrease toward zero, as the network increasingly assigns attention to self-loops, causing the embeddings to be dominated by each node’s own features rather than by relational information (see also \ref{appendix_2}). Importantly, the calibration graph is completely decoupled from the main target graph, ensuring that no information leaks between the two. 

\subsection{Ensemble clustering}
\label{subsec:ensemble}

After processing the graph with the GAT autoencoder, stars that are connected in the IoM space and are chemically similar receive higher attention weights, and their coherence is  enhanced in the reconstructed chemical space. Before directly identifying these groups, we accounted for the stability and reproducibility of our method. We note that two main factors introduce variability: first, each reconstructed space depends critically on the epoch at which the embeddings are analyzed; second, the choice of hyperparameters for the clustering algorithm can further alter the clustering outcome between runs. To ensure robust and reproducible findings, we followed the recommendations of \citet{unsupervised} and designed a clustering ensemble pipeline that uses OPTICS \citep{optics} as the base algorithm, quantifying both per-cluster and per-instance clustering stability. \par
We trained 100 GATs, first filtering out models with high reconstruction loss (only selecting those with \texttt{loss < 0.70}), and then selected the ten models with the largest area covered by the calibration edges. This ensures that the chosen models emphasize attention to chemically similar stars while maintaining accurate reconstruction of the chemical space. \par

 OPTICS identifies clusters by estimating the reachability distance, which quantifies how far a point is from its nearest dense region in the dataset. The algorithm can be tuned through three main hyperparameters: \texttt{min\_cluster}, which specifies the minimum number of points required to define a cluster; \texttt{xi}, which controls the sensitivity of the algorithm to changes in density; and \texttt{min\_samples}, which determines the minimum number of points in the neighborhood of a data point for it to qualify as a core point. Among these, \texttt{min\_samples} has a particularly strong influence on the shape of the reachability plot: if set too high, meaningful substructures may be merged or overlooked, whereas if set too low, the algorithm may detect spurious or noisy clusters. The hyperparameters used in each step are presented in Table \ref{tab:optics_params}. Additionally, before clustering the data were standardized using the \texttt{StandardScaler} from \texttt{scikit-learn} \citep{scikit-learn}, and the reachability distances were calculated using the \texttt{Manhattan} metric as in \citet{Spina_2022}. Moreover, clustering algorithms can struggle when the dataset contains groups of vastly different sizes. In particular, our reconstructed spaces contained one cluster that was much larger than all the others (see Figure~ \ref{fig:optics}), making it difficult to extract all clusters using a single set of hyperparameters. To address this, we employed a two-step approach: first, we used a set of hyperparameters to split the space into two broad groups; then, after extracting this large group, we analyzed the remaining stars separately to resolve the smaller substructures. Astrophysically, this corresponds to first separating accreted from in-situ stars, and then resolving the accreted substructure, which has a smaller cluster signal. The ensemble clustering technique across ten reconstructed spaces operates as follows:
\begin{enumerate}
    \item In each clustering run, we set the hyperparameters to partition the dataset into two broad groups: the largest cluster observed, and a second cluster containing the remaining substructure. We then extracted this large cluster in each run and computed, for each star, a membership probability defined as
    \begin{equation}
    p_{\rm mem} = \frac{\text{Number of times assigned to a cluster}}{\text{Number of runs}}.
    \label{eq:mem}
    \end{equation}
    Stars with a probability $p_{\rm mem} \geq 0.8$ were assigned to the cluster and removed from the analysis.
    \item We re-ran OPTICS on each of the ten reconstructed spaces for the remaining stars using the hyperparameters listed in the second column of Table \ref{tab:optics_params}. These ranges were selected to maximize the recovery of GCs in our sample in terms of both homogeneity ($H$) and completeness ($C$) parameters. To identify clusters that remain consistent across runs, we compared all pairs of clusters (on each star's unique identifier) from different runs using the Jaccard similarity index:
    \begin{equation}
    J = \frac{|A \cap B|}{|A \cup B|}
    \label{eq:jaccard}
    \end{equation}
    Clusters with $J \geq 0.5$ were considered matching, and those that matched in at least five runs across the ensemble were flagged as stable clusters. The threshold $J \geq 0.5$ enforces minimal overlap between clusters; for clusters of equal size, this corresponds to sharing at two thirds of the members. While this threshold may seem lenient, low-probability or spurious assignments can later be filtered by considering only stars with high membership probabilities.

   \item We built a graph where each node represents a stable cluster from an individual run (e.g., cluster 3 from run 0), and nodes were connected if their Jaccard similarity was $\geq$ 0.5. This approach handles transitivity: if cluster 3 from run 0 overlaps with cluster 5 from run 1, and cluster 5 from run 1 overlaps with cluster 4 from run 2, then clusters 3, 5, and 4 will merge into the same consensus cluster, even if the direct overlap between clusters 3 and 4 is below the threshold. Each connected component of this graph became a single consensus cluster.

   \item For each consensus cluster, we assigned to each star a probability of membership in the cluster, as done in Step 1 (equation \ref{eq:mem}). Each star was then assigned to the cluster for which it has the highest membership probability.
    
\end{enumerate}

\begin{table}[h!]
\centering
\begin{tabular}{lcc}
\hline
Hyperparameter & Step 1  & Step 2  \\
\hline
\texttt{min\_samples} & 100 & [28, 40] \\
\texttt{min\_cluster} & 700 & 30 \\
\texttt{xi} & 0.001 & 0.001 \\
\hline
\end{tabular}
\caption{OPTICS hyperparameters used in the two-step clustering procedure.}
\label{tab:optics_params}
\end{table}

This approach ensures robust and reproducible results, recovering only structures that are consistently present across all autoencoder runs. At the same time, it accommodates clusters of different sizes, allows the assignment of soft membership probabilities to each star, and enables assessment of the overall stability of the method.

\section{Supplementary analyses}

\subsection{Learning curve}
\label{appendix_2}
\begin{figure}[h!]
    \centering
    \includegraphics[width=\columnwidth]{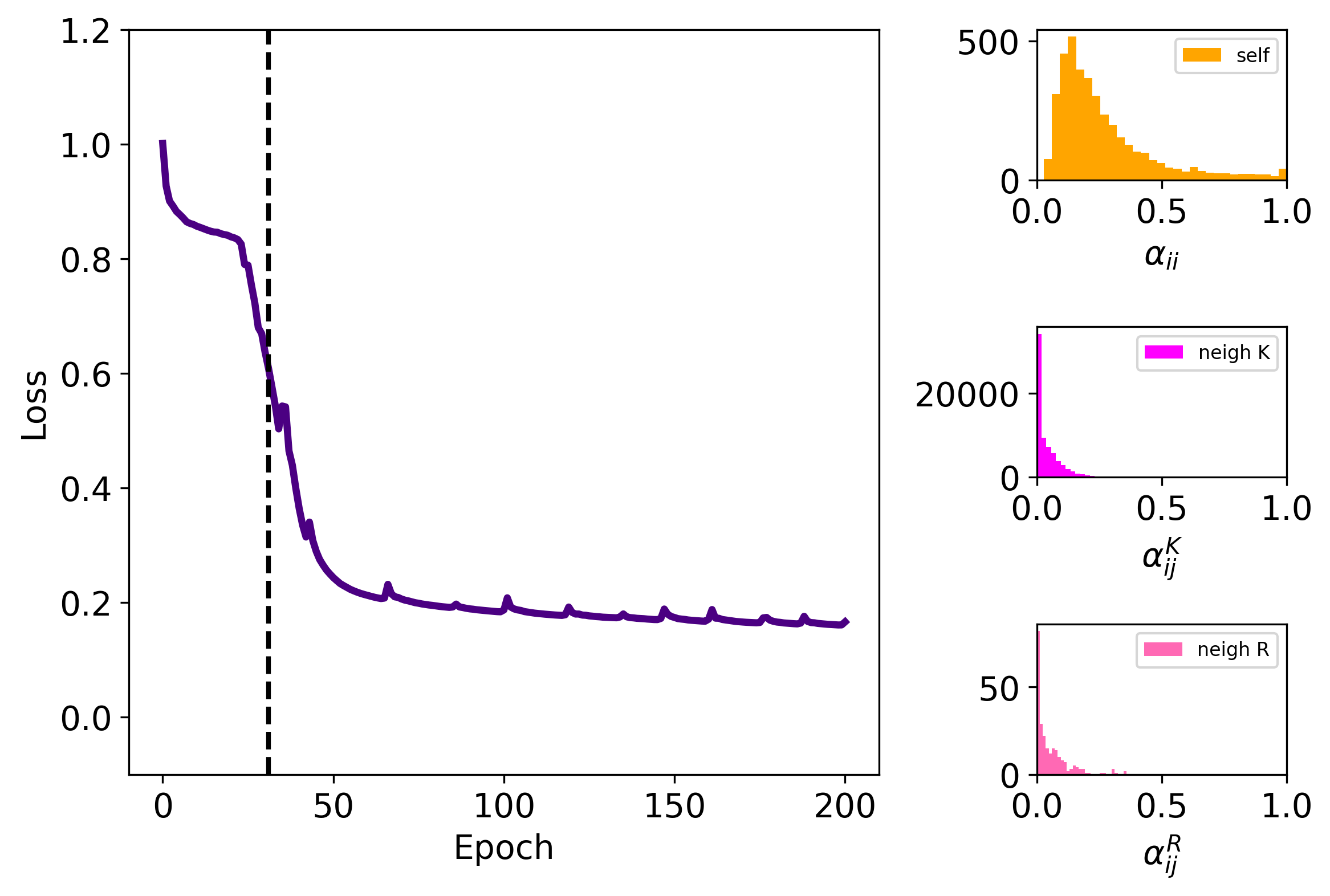}
    \caption{Learning curve over the 200 epochs. The vertical dashed line shows the epoch at which the reconstructed space in analyzed. The lateral panels show the self, kinematics defined, and random inter-cluster edge weights. }
    \label{fig:loss}
\end{figure}

We show a representative learning curve for one of the 100 trained GAT autoencoders in Figure~\ref{fig:loss}. The side panels display the distributions of the different attention weights: self-loops ($\alpha_{ii}$, top), kinematically defined edges ($\alpha^{k}_{ij}$, middle), and random calibration edges ($\alpha^{r}_{ij}$, bottom).

The distributions of the two inter-star attention weights, $\alpha^{k}_{ij}$ and $\alpha^{r}_{ij}$, show that a substantial fraction of links have been down-weighted or effectively "broken", though not all. In contrast, the distribution of self-loop weights (shown in orange) indicates that these connections have not yet come to dominate the training, as their values remain mostly between 0 and 0.5. If we were to repeat this analysis at epoch 200, all self-loop attention weights would approach 1 while all other weights would tend toward 0—signaling that the GAT has effectively collapsed into a standard feed-forward NN.

\subsection{Further tests to the GSE dichotomy}
\label{appendix_3}
\begin{figure}[h!]
    \centering
    \includegraphics[scale = 0.5]{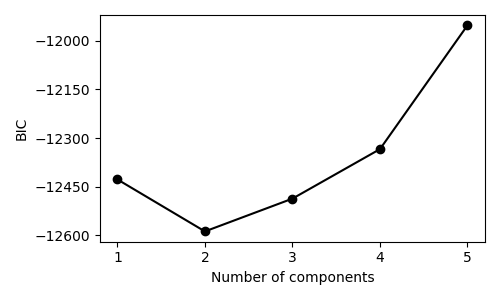}
    \caption{GMM fit to all GSE stars simultaneously, using full covariance matrices, implemented with \texttt{sklearn.mixture.GaussianMixture}}
    \label{fig:gmms}
\end{figure}

A very commonly employed method to disentangle different stellar populations in high-dimensional chemical space is GMMs. If our two GSE populations were coming from a true bimodality, a GMM should prefer 2 components over one, or more. In Figure~ \ref{fig:gmms} we show how the Bayesian Information Criterion is minimized exactly at 2 components when fitting GSE 1 and 2 together. 

\begin{figure}[h!]
    \centering
    \includegraphics[width=\columnwidth]{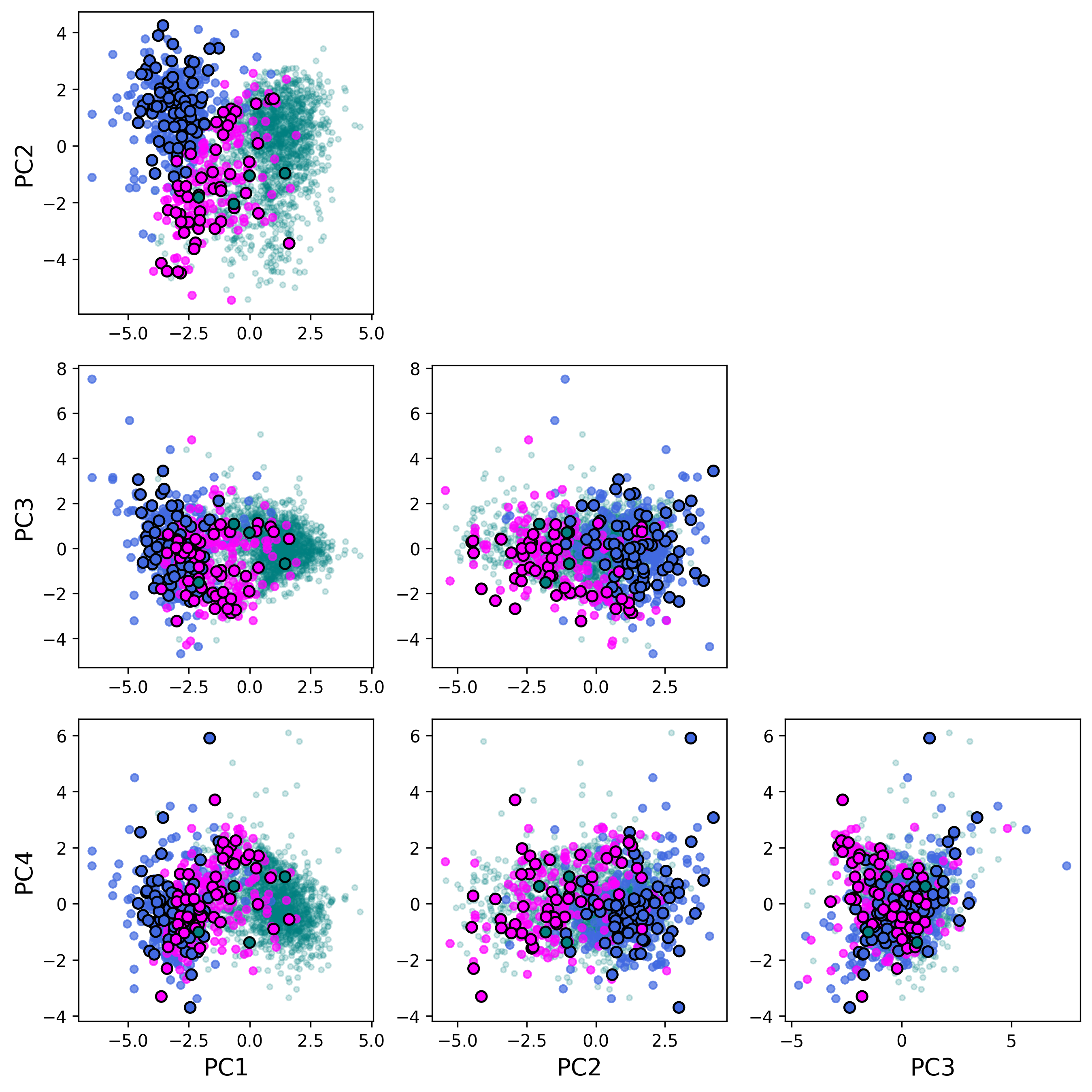}
    \caption{PCA projections onto first fours PCs and including the in-situ stars. As in Figure~ \ref{fig:pca}, stars from \citetalias{Dodd_2023} have a black edge.}
    \label{fig:pca_all}
\end{figure}

Finally, for completeness, we repeat the PCA reduction, this time onto four PCs and including the in-situ cluster versus both GSE clusters in Figure~ \ref{fig:pca_all} together with the loadings in Table \ref{tab:pca_loadings}.

\begin{figure*}[!t]
    \centering
    \includegraphics[width=0.8\textwidth]{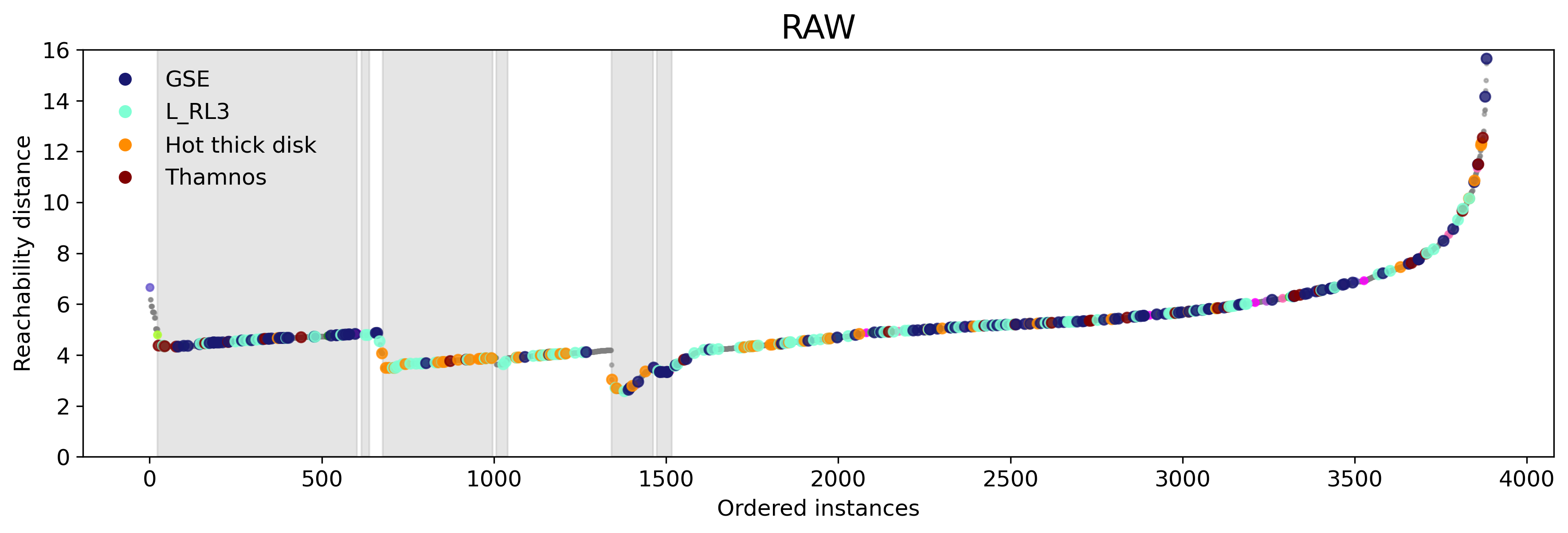}\\[1ex]
    \includegraphics[width=0.8\textwidth]{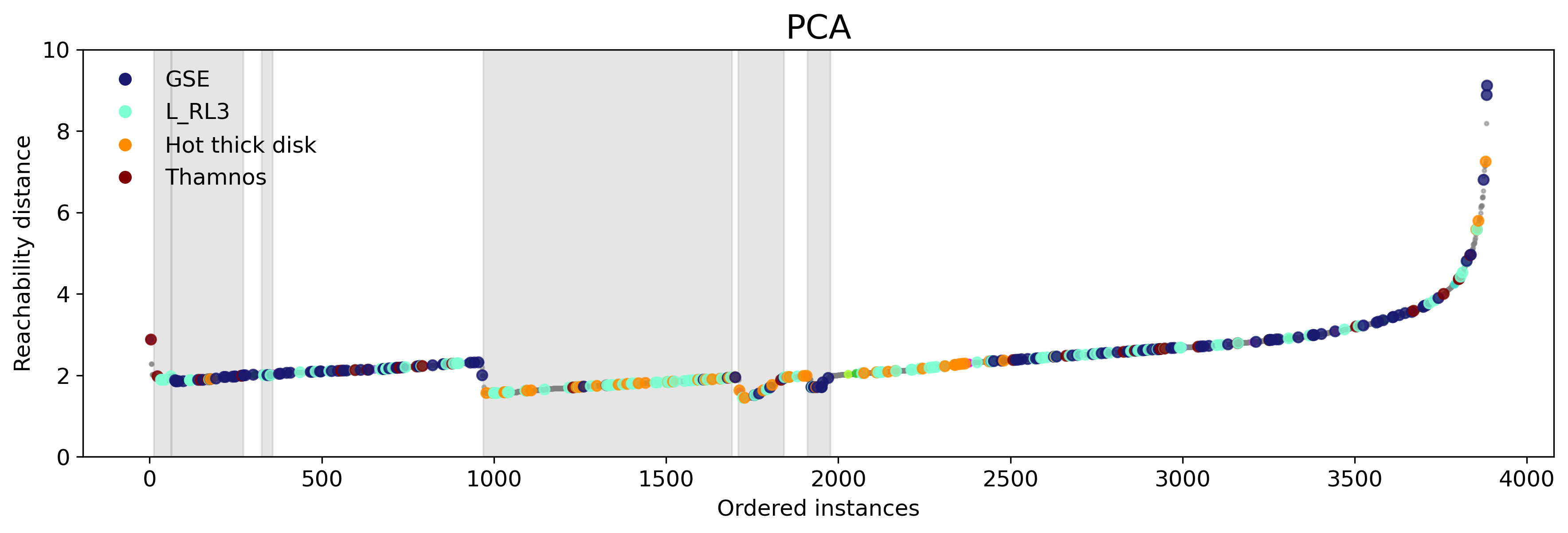}
    \includegraphics[width=0.8\textwidth]{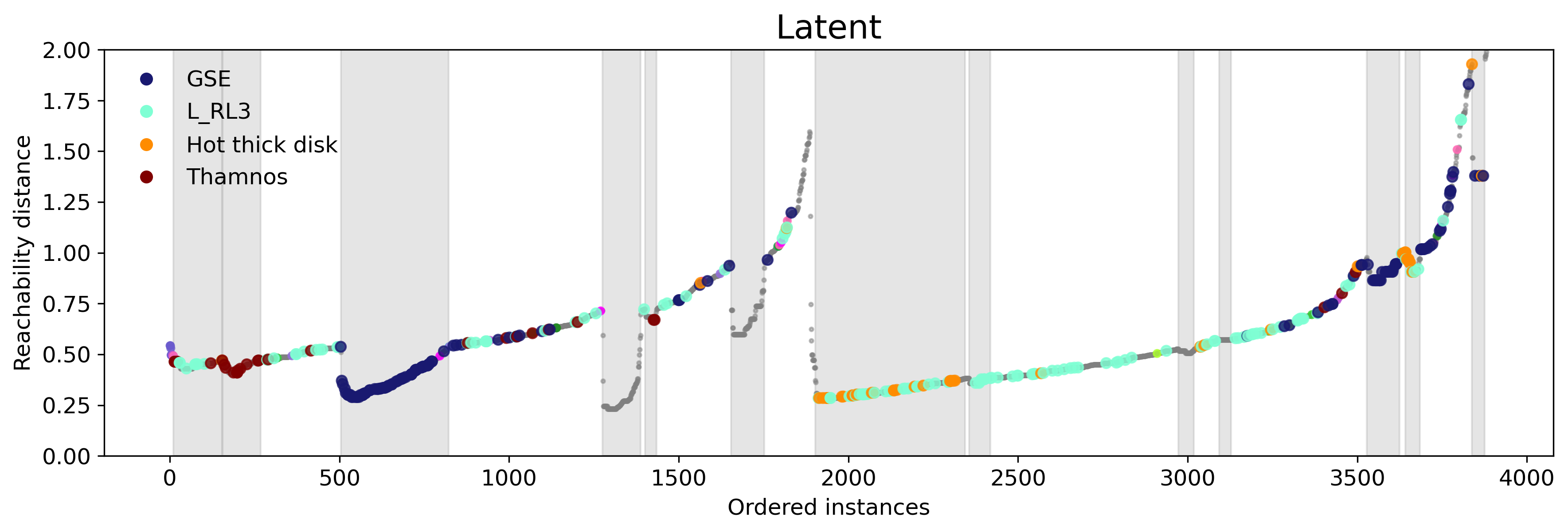}
    \caption{OPTICS run on raw chemical abundance space (top), PCA reduced to 6 components (middle), and the latent space of the same model shown in Figure~\ref{fig:optics}(bottom). All three runs have been run with \texttt{min\_samples} = 38, \texttt{xi}=0.0001 and \texttt{min\_cluster}=30}.
    \label{fig:raw_pca}
\end{figure*}

\begin{table}[htbp]
\centering
\begin{tabular}{lcccc}
\hline
Element & PC1 & PC2 & PC3 & PC4 \\
\hline
$[$Fe/H$]$  & 0.382 & -0.162 & -0.140 & -0.030 \\
$[$Na/Fe$]$ & 0.296 & 0.300  & 0.110  & 0.032 \\
$[$Mg/Fe$]$ & 0.180 & -0.357 & 0.313  & -0.428 \\
$[$Ca/Fe$]$ & 0.125 & -0.427 & -0.066 & 0.275 \\
$[$Nd/Fe$]$ & 0.193 & -0.356 & -0.404 & -0.044 \\
$[$Ba/Fe$]$ & 0.267 & -0.174 & -0.311 & -0.205 \\
$[$Y/Fe$]$  & 0.292 & 0.337  & -0.141 & -0.168 \\
$[$Cu/Fe$]$ & 0.233 & -0.048 & 0.037  & 0.697 \\
$[$Zn/Fe$]$ & 0.350 & -0.049 & 0.450  & -0.175 \\
$[$Ti/Fe$]$ & 0.316 & -0.076 & 0.420  & 0.321 \\
$[$Sc/Fe$]$ & 0.331 & 0.284  & 0.115  & -0.180 \\
$[$V/Fe$]$  & 0.329 & -0.041 & -0.362 & -0.020 \\
$[$Mn/Fe$]$ & 0.172 & 0.460  & -0.249 & 0.119 \\
\hline
\hline
Explained Variance & 0.398 & 0.224 & 0.079 & 0.069 \\
\hline
\end{tabular}
\caption{PCA loadings for the first four principal components and their explained variance.}
\label{tab:pca_loadings}
\end{table}

PC1 primarily separates the in-situ stars from GSE. The positive loadings for most elements indicate that this axis reflects overall stellar enrichment, with MW stars forming in a more chemically enriched environment than GSE stars. PC2 mainly separates GSE 1 from GSE 2, and is dominated by [Mg/Fe], [Ca/Fe], [Nd/Fe] (negative loadings) versus [Mn/Fe], [Na/Fe], and [Y/Fe] (positive loadings). This combination contrasts $\alpha$- and r-process enrichment against odd-Z and Fe-peak elements, highlighting the chemical distinction between the metal-poor, $\alpha$-enhanced GSE 1 and the more metal-rich, chemically extended GSE 2. Notably, GSE 2 lies closer to the in-situ population along PC1, hinting a more evolved composition.

\subsection{OPTICS on raw, PCA reduced and latent spaces}
\label{appendix_1}

We show the OPTICS fit on the raw abundances and on PCA-reduced abundances in Figure~\ref{fig:raw_pca} (top and middle panels). The PCA is run with six components, which explain just over 80\% of the variance. While some clusters are detected—particularly in-situ stars—these spaces show much less overall structure, with no GCs recovered. We also apply OPTICS to the latent space of the same model shown in Figure~\ref{fig:optics}. Although the overall structure appears similar by eye, the reachability plot shows more pronounced dips when clustering in the reconstructed space (note the scale on the y axis is different in Figures~\ref{fig:optics} and \ref{fig:raw_pca}, bottom panel). This figure also illustrates the necessity of the two-step ensemble clustering approach: the large valley corresponding to the in-situ cluster is only partially recovered in a single run, whereas the other dips in the reachability plot are captured robustly. Additionally, this OPTICS run finds more clusters than the one shown in Figure~ \ref{fig:optics}; some correspond to reproducible, physically meaningful structures, while others appear to be spurious overdensities. The ensemble clustering pipeline allows us to disentangle these two types of clusters effectively.

\end{document}